\documentclass[]{aa}  

\usepackage{graphicx}
\usepackage{ulem}
\usepackage{hyperref}
\usepackage{txfonts}
\usepackage{todonotes}
\usepackage{caption}
\usepackage{subcaption}

\usepackage{titlesec}
\newcommand{\jozef}[1]{\textcolor{black}{#1}}
\newcommand{\referee}[1]{\textcolor{black}{#1}} 

\hypersetup{colorlinks=true, linkcolor=blue, citecolor = blue, urlcolor=cyan}

\begin{document} 


\title{Probing the two-body decaying dark matter scenario with weak lensing and the cosmic microwave background}
\titlerunning{Probing the two-body decaying dark matter scenario with WL and the CMB}
  \author{Jozef Bucko\thanks{jozef.bucko@uzh.ch}\inst{1}
          \and
          Sambit K. Giri\inst{1,2}
          \and
          Fabian Hervas Peters\inst{1,3}
          \and 
          Aurel Schneider\inst{1}
          }

  \institute{Institute for Computational Science, University of Zurich, Winterthurerstrasse 190, 8057 Zurich, Switzerland \and
  Nordita, KTH Royal Institute of Technology and Stockholm University, Hannes Alf\'vens v\"ag 12, SE-106 91 Stockholm, Sweden \and 
  Université Paris-Saclay, Université Paris Cité, CEA, CNRS, Astrophysique,
Instrumentation et Modélisation Paris-Saclay, 91191 Gif-sur-Yvette, France
             }
  \date{Received: XX, XX, XXXX. Accepted: YY, YY, YYYY, Report Number: NORDITA 2023-020}


 
  \abstract
   {
   Decaying dark matter (DDM) scenarios have recently regained attention due to their potential ability to resolve the well-known clustering (or $S_8$) tension between weak lensing (WL) and cosmic microwave background (CMB) measurements. In this paper, we investigate a well-established model where the original dark matter particle decays into a massless particle and a massive daughter particle. The latter obtains a velocity kick during the decay process that results in the suppression of the matter power spectrum at scales that are observable with WL shear observations. We perform the first fully non-linear WL analysis of this two-body decaying dark matter ($\Lambda$DDM) scenario, including intrinsic alignment and baryonic feedback processes. We used the cosmic shear band power spectra from \textit{KiDS-1000} data and combined it with temperature and polarisation data from \texttt{Planck} in order to constrain the $\Lambda$DDM model. We report new limits on the decay rate and mass splitting parameters that are significantly stronger than previous results, especially in the case of low-mass splittings. Regarding the $S_8$ tension, we found a reduction from about 3 to 2 $\sigma$, depending on which statistical measure is applied. We therefore conclude that the two-body $\Lambda$DDM model is able to reduce the $S_8$ tension without convincingly solving it. 
   }


   \maketitle
%

\section{Introduction}
\label{sec:intro}

The standard $\Lambda$-cold dark matter ($\Lambda$CDM) cosmology provides an outstanding description of the Universe, explaining a wide range of cosmic observations such as the cosmic microwave background (CMB), baryonic acoustic oscillations (BAO), large-scale structure formation, and Big Bang nucleosynthesis (BBN). Despite its tremendous success, there are still questions to which $\Lambda$CDM, as understood nowadays, cannot provide satisfying answers, including the nature of dark matter and dark energy. Moreover, with progressively precise measurements, several discrepancies within the model have emerged. 
An example of such a discrepancy is the mild yet persistent clustering amplitude tension between the CMB and weak lensing (WL) measurements, often expressed via the parameter $S_8=\sigma_8 \sqrt{\Omega_{\rm m}/0.3}$, with $\sigma_8$ and $\Omega_{\rm m}$ describing the clustering amplitude and the total matter abundance. More specifically, CMB measurements from the {\tt Planck} collaboration have yielded $S_8 = 0.834 \pm 0.016$ \citep{planck_2018},
while a variety of low-redshift surveys report consistently lower values. For example, the Kilo-Degree Survey \citep[{\tt KiDS};][]{Kuijken_2019_kids,Giblin_2021_kids_1000_catalogue,Hildebrandt_2021_kids1000_catalogue} obtained a value of $S_8 = 0.760^{+0.016}_{-0.038}$ \citep[]{kids_1000}, in agreement with (albeit slightly lower than) the results from the Hyper Supreme-Cam \citep[{\tt HSC};][]{aihara_2017_hsc,hsc_y1_weak_lensing} and the Dark Energy Survey \citep[{\tt DES};][]{DES_2005,amon_2022_des}.

It remains unclear whether the $S_8$~tension emerges from an insufficient modelling of the non-linear clustering of matter \citep[e.g.][]{Tan:2022wob,arico2023des} or the modelling of cosmic shear \citep{Garcia_Garcia_2021} or whether one has to look beyond the standard $\Lambda$CDM. Resolving the $S_8$~tension can be achieved by suppressing the matter power spectrum at scales $k\sim 0.1$-$1$~h/Mpc, which most substantially influence the clustering amplitude value $S_8$ \citep[e.g.][]{amon2022non}. Such suppression may be obtained by a number of extensions of $\Lambda$CDM, such as cold-warm dark matter \citep{Schneider:2019xpf,Parimbelli_2021}, cannibal dark matter \referee{\citep{cannibal_DM,Heimersheim_2020}}, models involving interactions between dark matter and dark radiation at early times \referee{\citep{ethos_original,ethos_simulations,Rubira_2023,Joseph_2023}}, scenarios introducing an interaction between dark matter and dark energy \citep[and references therien]{portsidou_2013_dm_de_, poulin_2022_drag} or baryons and dark energy \citep{ferlitto_2022}, or models assuming unstable dark matter particles \referee{\citep{Enqvist:2015ara, Enqvist:2019tsa,Murgia_2017, Abellan_2021,Chen_2023:DES_decays,CHOI2022136954,bucko_2022_1bddm}}.

The class of decaying dark matter models includes several different scenarios. In the simplest case, a fraction of dark matter decays into a relativistic component (often assumed to be dark radiation). However, such a model is strongly constrained by CMB observations \cite[see e.g.][]{Hubert_2021,Simon_2022,bucko_2022_1bddm}, as it affects the cosmic background evolution. An alternative and only somewhat more complex scenario assumes the initial dark matter particles decay into a pair made up of a massless and a massive particle, the latter obtaining a velocity kick during the decay process. This scenario is referred to as a two-body decaying dark matter model and is denoted as '$\Lambda$DDM' hereafter.

A direct consequence of the $\Lambda$DDM model is the free streaming process of the stable decay products, which alters the gravitational collapse of cosmic structures. This effect is relevant at scales set by the free-streaming length of the stable daughter particles and thus by the magnitude of the velocity kicks ($v_k$) they receive as a consequence of energy-momentum conservation. As a result, the matter power spectrum is suppressed during late times at scales above $k\sim 0.1$ h/Mpc \cite{Wang_Zentner_2012}.

In addition to the aforementioned reason, there are arguments motivated by particle physics to consider models in which dark matter is not stable over cosmic time. First of all, such a stability condition does not emerge naturally; it usually requires additional assumptions such as $Z_2$ symmetry \citep{HAMBYE:2011Wq}. Moreover, there are numerous theoretical models involving dark matter decays, such as sterile neutrinos \citep{sterile_neutrinos_1,Adhikari_2017}, R-parity violation  \citep{BEREZINSKY1991382,KIM200218}, or super weakly interacting massive particles \citep{superWIMPS1,superWIMPS2}.

The two-body decaying dark matter model ($\Lambda$DDM) has been studied from various angles over the past decade (e.g. by using perturbation theory \citep{Wang_Zentner_2012,Abellan_2021,Simon_2022} or $N$-body simulations ranging from individual galaxies \citep{Peter2010DarkmatterDA} to a large-scale structure \citep{Cheng_2015}). For example, \cite{Mau_2022} obtained constraints on the $\Lambda$DDM model based on Milky Way satellite counts, while \cite{Wang:2013rha} and \cite{Fuss_Garny_2022} used Lyman-$\alpha$ forest data to constrain the two-body decay rate in the regime of low-mass splittings. Additionally, \cite{Abellan_2021} and \cite{Simon_2022} considered {\tt Planck} CMB observations together with supernova type Ia (SNIa) data and BAO to derive constraints on two-body decays. After including priors from WL observations, they reported a reduction of the $S_8$ tension for a best-fitting $\Lambda$DDM model with $\tau=120$~Gyr and $v_k/c\simeq 1.2$\% \citep{Simon_2022}.  

In this work, we perform the first WL analysis of the $\Lambda$DDM model using cosmic shear data from the \textit{KiDS-1000} survey \citep{kids_1000}. The non-linear clustering predictions were thereby modelled using an emulator based on a suite of $N$-body simulations. Along with the WL analysis, we also performed a re-analysis of the {\it Planck 2018} CMB temperature and polarisation data as well as a combined WL plus CMB modelling investigating the potential of $\Lambda$DDM to solve the $S_8$ clustering tension.  
   
Our paper is organised as follows: In Section~\ref{sec:2bDDM}, we describe basic physics and implications of the $\Lambda$DDM model, while Section~\ref{sec:theory} provides a detailed overview of our modelling of WL and CMB observables. In Section~\ref{sec:mcmc}, we comment on choices made in relation to  model inference, and in Section~\ref{sec:results}, we describe our results and compare them to recent studies before concluding in Section~\ref{sec:conclusion}. 
In Appendix~\ref{app:compare_model}, we compare the results of our $N$-body simulations to previous studies. Appendix~\ref{app:cosmology_dependence} presents a discussion on the cosmology dependence of the $\Lambda$DDM effects, and Appendix~\ref{app:parameter_effects} details the effects two-body decays and baryons have on WL and CMB observables. In Appendix~\ref{app:tension}, we study more closely the tension between WL and CMB data in the context of the $\Lambda$DDM model, and finally,  Appendix~\ref{app:mcmc_results} provides more detailed information about the parameters we obtained from model inference.

\section{Two-body decaying dark matter}
\label{sec:2bDDM}
In the $\Lambda$DDM model, an original ('mother') dark matter particle decays into a slightly lighter, stable particle and a relativistic massless relic, while the energy released during the decay process is split between two product species. We  describe the basic theoretical properties of the $\Lambda$DDM model in Section~\ref{sec:theory_2bDDM}. In Section~\ref{sec:sim_2bDDM}, we discuss our $\Lambda$DDM $N$-body simulations, and finally, we introduce a new emulator of the $\Lambda$DDM non-linear matter power spectra in Section~\ref{sec:emul_2bDDM}.

\begin{figure*}[!ht]
    \centering
    \includegraphics{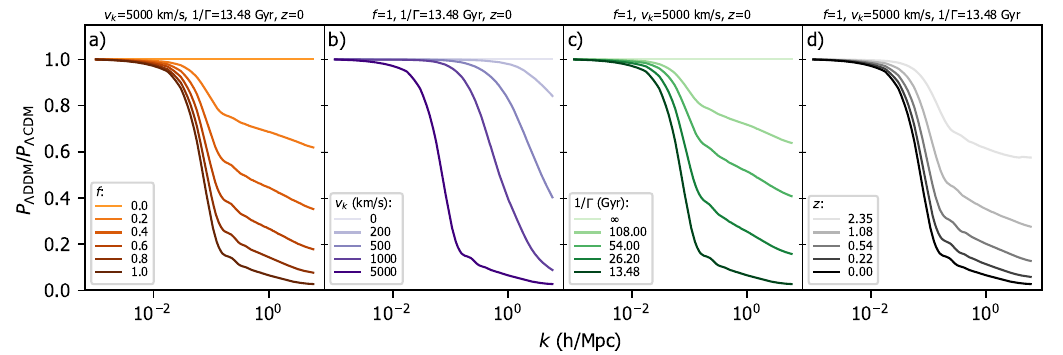}
    \caption{Effects of two-body decays on the non-linear matter spectrum from $N$-body simulations. The fiducial ($\Lambda$CDM) power spectrum corresponds to a constant line equal to one, while the power suppressions for varying $f$, $v_k$, $\Gamma$, and $z$ are shown in the panels (a), (b), (c), and (d).}
    \label{fig:sim_results_2body}
\end{figure*}

\subsection{Theory}\label{sec:theory_2bDDM}
The $\Lambda$DDM model is simple enough to be described by two phenomenological parameters. The first parameter is the decay rate $\Gamma$ controlling the frequency of the decay processes. The second parameter corresponds to the velocity kick $v_k$ of the massive daughter particle, which is directly linked to the mass ratio between the mother and daughter particles. Some authors replace the velocity kick magnitude $v_k$ with the ratio $\varepsilon$ of the rest-mass energy
\begin{eqnarray}
    \label{eq:epsilon_def}
    & &\varepsilon = \frac{1}{2}\left(1 - \frac{m^2_{\rm wdm}}{M^2_{\rm dcdm}}\right),
\end{eqnarray}
where $m_{\rm wdm}$ and $M_{\rm dcdm}$ denote the rest mass of the warm daughter and the decaying cold mother dark matter particles, respectively \citep[see e.g.][]{blackadder_Savvas_epsilon}. The momentum of the daughter particle in the centre-of-momentum frame is $p_{\rm wdm}=m_{\rm wdm}c\varepsilon/ \sqrt{1-2\varepsilon}$ \citep{Abellan_2021}, where $c$ is the speed of light. We note that in the non-relativistic limit, we obtained a simple relation $v_k = \varepsilon c$. 

The background evolution of the energy densities of both the cold (mother) and warm (daughter) dark matter species as well as the massless, dark radiation (daughter) component can be written as \citep{Wang_Zentner_2012}
\begin{eqnarray}
    \label{eq:rho_DDM}& &\dot{\rho}_{\rm dcdm} + 3\mathcal{H}\rho_{\rm dcdm} =-a \Gamma \rho_{\rm dcdm},\\
    \label{eq:rho_SDM}& &\dot{\rho}_{\rm wdm} + 3\mathcal{H}\rho_{\rm wdm} =a \Gamma \frac{M^2_{\rm dcdm}+m^2_{\rm wdm}}{2M^2_{\rm dcdm}}\rho_{\rm dcdm},\\
    \label{eq:rho_DR}& &\dot{\rho}_{\rm dr} + 4\mathcal{H}\rho_{\rm dr} =a \Gamma \frac{M^2_{\rm dcdm}-m^2_{\rm wdm}}{2M^2_{\rm dcdm}}\rho_{\rm dcdm},
\end{eqnarray}
or using $\varepsilon$ instead of particle masses, as
\begin{eqnarray}
    \label{eq:rho_DDM}& &\dot{\rho}_{\rm dcdm} + 3\mathcal{H}\rho_{\rm dcdm} =-a \Gamma \rho_{\rm dcdm},\\
    \label{eq:rho_SDM}& &\dot{\rho}_{\rm wdm} + 3\mathcal{H}\rho_{\rm wdm} =a \Gamma(1-\varepsilon) \rho_{\rm dcdm},\\
    \label{eq:rho_DR}& &\dot{\rho}_{\rm dr} + 4\mathcal{H}\rho_{\rm dr} =a \Gamma \varepsilon \rho_{\rm dcdm}.
\end{eqnarray}
In the above equations, $\mathcal{H}$ is the conformal Hubble parameter, and $\rho_{i}$ is the energy density of species $i$. The subscripts 'dcdm', 'wdm', and 'dr' refer to the cold, warm, and massless species. Dots denote derivatives with respect to \referee{conformal} time, and $a$ stands for the scale factor.

Next to the two model parameters $\Gamma$ and $v_k$, we include a third parameter
\begin{eqnarray}\label{fraction}
    \label{eq:f_definition}
    & &f=\frac{\Omega_{\rm dcdm}}{\Omega_{\rm dcdm}+ \Omega_{\rm cdm}},
\end{eqnarray}
that allows for a scenario where only a faction $f$ of the total initial dark matter fluid is unstable (while the remaining dark matter corresponds to a stable CDM particle). Here, we have introduced the abundances of the stable ($\Omega_{\rm cdm}$) and unstable ($\Omega_{\rm dcdm}$) dark matter species, respectively. With the above description, one can in principle study the full parameter space of two-body decays by taking arbitrary $\Gamma$, $\varepsilon$, and $f$ values, with limiting cases $\Gamma \rightarrow 0$, $\varepsilon \rightarrow 0$ or $f \rightarrow 0$ approaching the $\Lambda$CDM cosmology and $\varepsilon \rightarrow 1/2$ approaching one-body decays. However, as recent studies have demonstrated \citep[and references therein]{Wang_Zentner_2012,Mau_2022}, $\Lambda$DDM models with a very large decay rate and velocity kicks are ruled out by observations, as they lead to a strong power suppression at $k< 0.1$~h/Mpc. We therefore focus on the regime with late-time decays ($\Gamma \lesssim H_0$) and non-relativistic velocity kicks ($v_k \ll c$) throughout this paper.

\subsection{Simulations}\label{sec:sim_2bDDM}
Considering only non-relativistic decays,  we implemented the $\Lambda$DDM model into the $N$-body code \texttt{PKDGRAV3} \citep{Potter_2017_pkdgrav3}, a tree-based gravity solver based on fast multi-pole expansion and adaptive time stepping. Following the theoretical description \eqref{eq:rho_DDM}-\eqref{eq:rho_DR}, we found that at first order, the background equations remain unmodified. With this approximation, there is no energy transfer between the radiation and dark matter components caused by the decay process. Therefore, we kept the background cosmology implementation of \path{PKDGRAV3} unchanged and implemented only the non-relativistic velocity kicks received by the WDM particles. We note that this differs from the one-body DDM model studied in \cite{bucko_2022_1bddm}, where the background evolution had to be modified. 

The two-body decays are implemented into \path{PKDGRAV3} via a function {\tt pkdDecayMass}, which is revisited at each global integration time step (separated by a time interval $\Delta t$). The decay probability of a given (not yet decayed) particle at time step $i$ is $P = \Gamma \Delta t$. Thus, a number of $\Delta N^i_{\rm wdm} = \Gamma \Delta t N^i_{\rm dcdm}$ simulation particles undergo the decay process, where $N^i_{\rm dcdm}$ denotes the number of unstable DCDM particles at time step $i$. The particles that are about to decay are chosen randomly from all CDM particles. Immediately after the decay, they obtain a uniform velocity kick in a random direction. Importantly, these particles are flagged and added to a set of already decayed particles that are excluded from the decay process occurring in future time steps. 

In Fig.~\ref{fig:sim_results_2body}, we plot the ratios of simulated $\Lambda$DDM to $\Lambda$CDM power spectra for varying values of $f$, $v_k$, and $1/\Gamma$ (see panel a, b, and c). In general, the two-body decaying dark matter model leads to a suppression of the total matter power spectrum at small scales and leaves the large scales unchanged. The amplitude of the suppression, as well as the scale of the downturn, depends on the values of the $\Lambda$DDM parameters. The fraction of decaying dark matter as well as the decay rate both affect the amplitude of the suppression, while the value of the velocity kick primarily influences the position of the downturn along the $k$-axis. The latter can be understood by the fact that larger streaming velocities are able to affect the formation of structures at larger scales.

The redshift dependence of the power suppression is shown in panel (d) of Fig.~\ref{fig:sim_results_2body}. Not surprisingly, the amplitude of the suppression increases towards lower redshifts. This behaviour is caused by the fact that more particles decay with time, causing a reduction in the clustering process compared to $\Lambda$CDM.

We ran a suite of $N$-body simulations for decay rates $\Gamma < 1/13.5$~Gyr$^{-1}$ and velocity kicks $v_k/c < 0.02$. All of our simulations were run assuming a fiducial cosmology with parameters $h_0 = 0.678,\, \Omega_{\rm m,0} = 0.307,\, \Omega_{\Lambda,0} = 0.693,\, \Omega_{\rm b,0} = 0.048,\, n_s = 0.966$, and $\sigma_8 = 0.883$. We obtained converged results at the scales $k\sim 0.01-10$ h/Mpc for box sizes of $L_{\rm box} = 125,250,512$~Mpc/h and particle numbers of $N = 256^3, 512^3, 1024^3$ (depending on the specific $\Lambda$DDM configuration). 
We compared the output of our simulations to results from previous works and found a good level of agreement (see Appendix~\ref{app:compare_model}). 
As we are primarily interested in the ratio of the non-linear power spectra between the $\Lambda$DDM and $\Lambda$CDM models, \jozef{one can expect that }the cosmology dependence is factored out to a large extent. We {tested to see whether} \jozef{verified that } the impact of the cosmology is much smaller than the suppression due to the two-body decays \jozef{(see Appendix~\ref{app:cosmology_dependence})}.

\subsection{Emulating the impact of dark matter decays}
\label{sec:emul_2bDDM}

\begin{figure*}
    \centering
    \includegraphics[width = 0.9\textwidth]{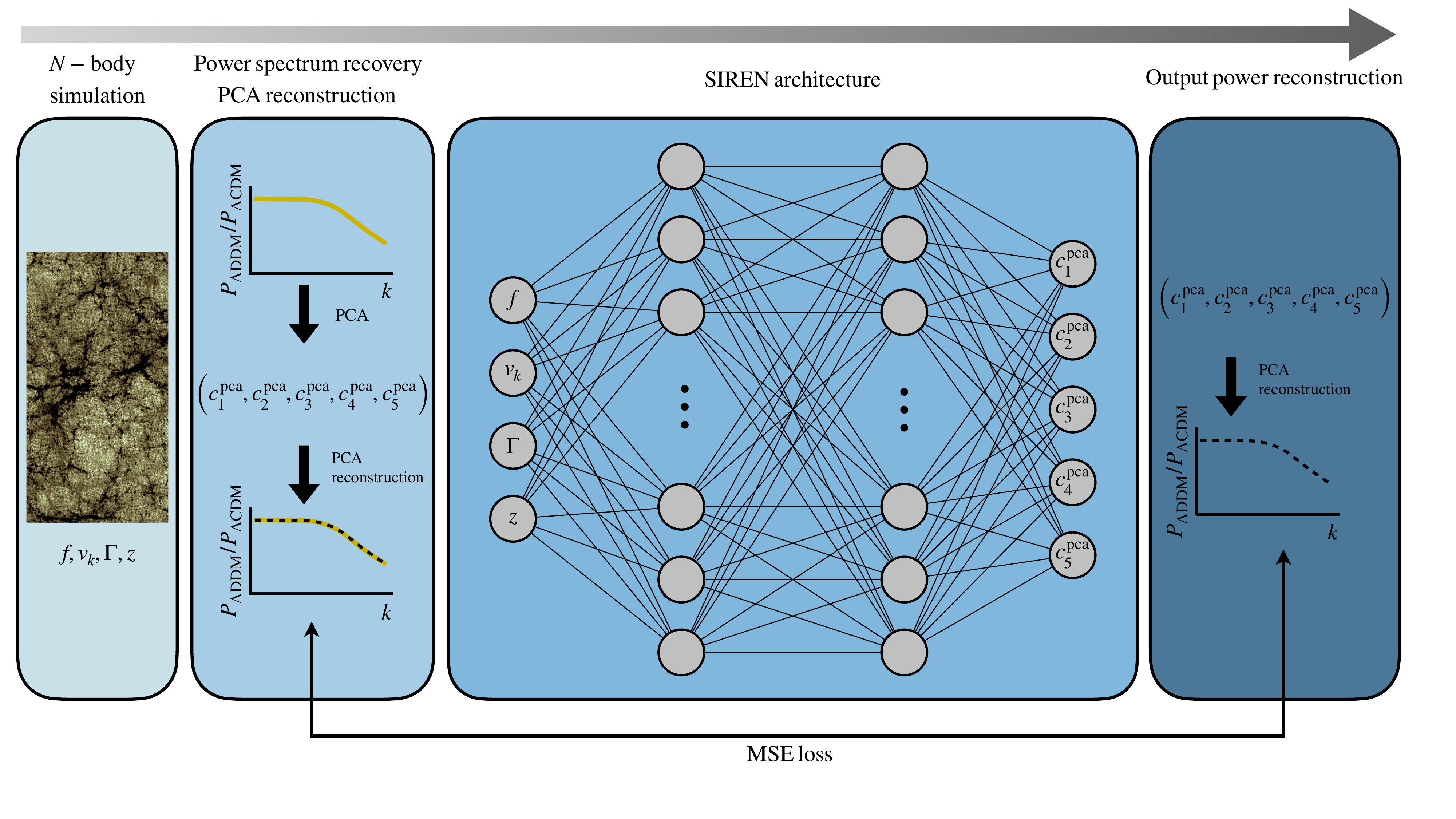}
    \caption{Flowchart describing the emulation process (from left to right). The $N$-body simulations first need to be run to provide the matter density distribution for the given DDM parameters and redshift $(f,v_k,\Gamma,z)$. Next, we extracted power spectra and calculated the resulting power suppression $P_{\rm DDM}/P_{\Lambda \rm{CDM}}$. Additionally, we performed PCA on $P_{\rm DDM}/P_{\Lambda \rm{CDM}}$ data, and using five principal components  $(c_1^{\rm pca},c_2^{\rm pca},c_3^{\rm pca},c_4^{\rm pca},c_5^{\rm pca})$, we recovered the original power suppression. Once the training data set was prepared, we trained the SIREN-like neural network. We used $(f,v_k,\Gamma,z)$ as the network input and output five PCA components. The PCA components obtained were used to reconstruct the power spectrum suppressions, which were then compared to the input ones. During the training process, the MSE loss between the input and output power suppression curves was minimised.}
    \label{fig:emulator_scheme}
\end{figure*}

In order to carry out a Bayesian inference analysis (Section~\ref{sec:mcmc}), we needed a fast modelling framework to explore the vast parameter space of astrophysics, cosmology, and dark matter models. As $N$-body simulations are not fast enough for this purpose, we built an emulator to account for the different $\Lambda$DDM parameters. 
The basic characteristics of our emulator-building procedure are shown in the flowchart of Fig.~\ref{fig:emulator_scheme}. First, we ran $\sim$100 gravity-only simulations for different dark matter parameters ($\Gamma$, $v_k$, and $f$) and measured the non-linear matter power spectra up to $k\sim 6\,h$/Mpc between $z=2.35$ and $z=0$. In the next step, we performed a principal component analysis \citep[PCA, see e.g.][]{hands-on-ML_2019} on the ratios of $\Lambda$DDM and $\Lambda$CDM matter power spectra $\mathcal{S}_{\Lambda\rm DDM}^{\Gamma,v_k,f}(k,z)$. We found that five PCA components are sufficient to describe the ratio of spectra with a reconstruction error of approximately $0.1$\%.

Next, we trained a neural network to model these five PCA components of the simulated power spectra ratios for a given parameter vector ($\Gamma,v_k,f,z$). The network output was then transformed back from the PCA representation to the power spectra ratios before being compared to the original (simulated) ratios used for the training and testing of the emulator. During the network training process, we minimised the differences between the predicted and true power spectra ratios in the training set. We considered the mean squared error (MSE) metric to quantify the differences.

We chose the architecture of the sinusoidal representation networks (SIRENs; \citealt{sitzmann2019siren}) for building our $\Lambda$DDM emulator. The SIRENs have been successfully shown to have good interpolation and signal reconstruction properties \citep{sitzmann2019siren}. The main difference compared to standard feed-forward architectures involves replacing the commonly assumed {\it ReLU} activation function with a sine function. We used the architecture with two hidden layers that each had 1024 neurons to perform the emulation task. During the training, we optimised the MSE with the Adam optimiser \citep{adam_optimizer}. To further fine-tune the SIREN architecture, we performed a hyperparameter optimisation for the network's learning rate $l_r$ and regularisation strength $\lambda$ using the Bayesian Optimization and Hyperband (BOHB) method \citep{bohb}. The entire training was performed using the \texttt{PyTorch} \citep{NEURIPS2019_9015} machine learning framework.

\begin{figure}
    \centering
    \includegraphics[width=\columnwidth]{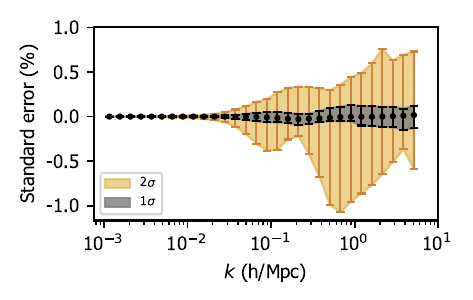}
    \caption{
    \jozef{Performance of the $\Lambda$DDM emulator at different scales.} The precision of the two-body decaying dark matter emulation procedure was assessed by testing the emulator on the data that was not used during training. The average prediction error was calculated for each of the 30 $k$-bins, and the 1$\sigma$ (grey) and 2$\sigma$ (golden) error bars of the prediction are highlighted.
    }
    \label{fig:emulator_performance}
\end{figure}

After training, we tested the emulator on separate data (i.e. the test set) and monitored the prediction mismatch in each of the 30 $k$-bins. We show the emulation performance in Fig.~\ref{fig:emulator_performance} and demonstrate that both 1$\sigma$ and 2$\sigma$ errors stay below 1\%. Our emulator thus efficiently predicts the response of decays on the non-linear matter power spectrum $\mathcal{S}_{\Lambda\rm DDM}^{\Gamma,v_k,f}(k,z)$. 
The prediction time of our emulator is a few milliseconds.
With this tool, we could model the non-linear power spectrum in the presence of dark matter decays as
\begin{equation}\label{nonlinPS}
P^{\rm nonlin}_{ \Lambda\rm DDM}(k,z) = \mathcal{S}_{\Lambda\rm DDM}^{\Gamma,v_k,f}(k,z) \times P^{\rm nonlin}_{\rm \Lambda CDM}(k,z),
\end{equation}
where $P^{\rm nonlin}_{\rm \Lambda CDM}(k,z)$ is obtained by multiplying the non-linear power spectrum from the {\tt revised\_halofit} method \citep{Takahashi_2012_revised_halofit}. Our emulator to model the non-linear $\Lambda$DDM power spectrum is published as part of the publicly available code {\tt DMemu}\footnote{\url{https://github.com/jbucko/DMemu}}.

\section{Data sets and modelling framework}
\label{sec:theory}

Here we first describe the observational  data that we use to study the $\Lambda$DDM  model. Later, in Section~\ref{sec:shear_modelling} and \ref{sec:cmb_modelling}, we present our modelling framework.

\subsection{Data sets}
Galaxy WL is a particularly promising observable to probe decaying dark matter models, as such scenarios tend to affect structure formation at late times and small cosmological scales. However, while primarily focusing on the WL analysis, we also include CMB data in our analysis. 

\jozef{In our study, we used the WL cosmic shear data of the {\it KiDS-1000} data release, obtained in five redshift bins between $0.1\lesssim z  \lesssim 1.2$ \citep{kids_1000}. We used the band power angular spectra measured at scales $118 \leq l \leq 1266$. Additionally, we included the {\it Planck 2018} high-$\ell$ power spectra  ($\ell\geq 30$) of temperature (TT), polarisation (EE), and their cross (TE) obtained from \cite{planck_V_power_spectra_likelihoods}.}
In the following subsections, we discuss how we {predicted} \jozef{computed these observational quantities} in our modelling framework.

\subsection{Cosmic shear modelling for {\it KiDS-1000}}
\label{sec:shear_modelling}
The {\it KiDS-1000} catalogue provides information about the shear of over 20 million galaxies, divided into five tomographic bins between $z\sim0.1$ and $z\sim 1.2$. The catalogue provides the basis of the auto- and cross-correlation band power of all five redshift bins. With eight data points for each spectrum, the {\it KiDS-1000} band power contains a total of 120 observational data points with correlated errors from the corresponding covariance matrix.

\subsubsection{Non-linear matter power spectrum}
We used the public version of the Boltzmann Solver {\tt CLASS} \citep{Blas_2011_CLASS} to calculate the $\Lambda\rm CDM$ matter power spectrum for any given set of cosmological parameters. For the non-linear modelling, we relied on the \texttt{revised halofit} method \citep{Takahashi_2012_revised_halofit} implemented in \texttt{CLASS}. Following \cite{planck_2018}, we assumed a single massive neutrino species with a fixed mass $m_\nu = 0.06$~eV throughout this work.

The process of baryonic feedback causes gas to be expelled out of galaxies and clusters, leading to a suppression of the matter power spectrum at small cosmological scales \citep[e.g.][]{vanDaalen:2011xb,Schneider:2015wta,Chisari:2019tus}. We note that this suppression is similar in shape to the one caused by dark matter decays \citep{Hubert_2021}, making it particularly important to include baryonic feedback in our modelling pipeline. We used \texttt{BCemu}\footnote{The \texttt{BCemu} code can be found at \url{https://github.com/sambit-giri/BCemu}.} \citep{Giri_2021_bcemu}, an emulation tool providing the suppression $\mathcal{S_{\rm b}}(k,z)$ due to baryonic feedback described in \cite{Schneider:2018pfw}. The term $\mathcal{S_{\rm b}}$ is a function of seven baryonic parameters and one cosmological parameter, namely, the ratio of the baryonic and total matter abundance.
However, we used the reduced three-parameter model presented in \cite{Giri_2021_bcemu}, where four parameters are fixed based on the results from hydrodynamical simulations. The three-parameter model consists of two parameters describing the gas distribution (log$_{\rm 10}M_c$, $\theta_{\rm ej}$) and one parameter describing the stellar mass ($\eta_\delta$) around galaxies and clusters. 
We refer to \cite{Giri_2021_bcemu} for detailed description of the model.

Finally, the response function from the two-body dark matter decays was multiplied with the non-linear, baryonified power spectrum, as shown in Eq.~\ref{nonlinPS}. {It} \jozef{The former}  was obtained from the emulator described in Section~\ref{sec:emul_2bDDM}. This modular method assumes that all dependence on cosmological parameters is captured by the power spectrum from the {\tt revised halofit} \jozef{(see Appendix~\ref{app:cosmology_dependence})}. At the same time, the responses from the baryonification and the two-body decays remain independent of cosmology. Regarding the baryonification method, this assumption has been validated before \citep{Schneider:2019xpf,Schneider:2019snl}. {In Appendix~\ref{app:cosmology_dependence}, we validated this assumption for the $\Lambda$DDM model.}

\subsubsection{Intrinsic alignment}
The effects from intrinsic galaxy alignments were modelled assuming the non-linear alignment (NLA) model as described in \cite{hildebrandt_2016_kids_450} and first published in \cite{Hirata_Seljak_2004}. The intrinsic alignment enters the band power modelling via a window function \citep[see eq.~9 in][]{bucko_2022_1bddm} that goes into galaxy-intrinsic and intrinsic-intrinsic terms of the cosmic shear angular power spectrum. Among the two intrinsic alignment parameters $A_{\rm IA}$ and $\eta_{\rm IA}$, we fixed the latter one to zero following the approach used in the standard {\it KiDS-1000} analysis \citep{kids_1000}.

\subsubsection{Angular power spectrum}

We investigated the impact of decaying dark matter on structure formation through the analysis of the cosmic shear angular power spectrum, including both auto-correlation and cross-correlation power spectra between different galaxy populations (tomographic bins). Our modelling approach follows the methodology presented in \cite{bucko_2022_1bddm}, and we refer the reader to that work for a more comprehensive description.

We used the multi-purpose cosmology calculation tool \texttt{PyCosmo} \citep{pycosmo} to model the cosmic shear angular power spectrum. The angular power was then converted into band powers following \cite{joachimi_2020}. In Fig.~\ref{fig:cls_modif_from_2bddm}, we show all band power coefficients together with the respective error bars.  The auto- and cross-band power measurements are illustrated in five \referee{\it KiDS-1000} redshift bins\referee{, always for multipoles} between $l \simeq 118$ to $l \simeq 1266$.

\subsection{CMB modelling}
\label{sec:cmb_modelling}

The Boltzmann Solver {\tt CLASS} can also be used for theoretical modelling of the CMB temperature and polarisation data.
Since the work of \cite{Abellan_2021}, {\tt CLASS} comes with an implementation of the $\Lambda$DDM model\referee{ in which the authors introduced a fluid approximation that reduces the computational costs significantly,} and we use this model throughout our work.\footnote{\url{https://github.com/PoulinV/class_decays}} To investigate the effects of $\Lambda$DDM cosmology on CMB data, we modelled the temperature and polarisation power spectra from the {\it Planck 2018} data release \citep{planck_V_power_spectra_likelihoods}. We adopted the same methodology as the one introduced in \cite{bucko_2022_1bddm} and refer the reader to that work for more details. We note, however, that our pipeline is tested for the $\Lambda$CDM cosmology reaching an excellent agreement with the results from \cite{planck_2018}. A comparison can be found in appendix~A of \cite{bucko_2022_1bddm}.

\section{Model inference}
\label{sec:mcmc}

We performed a number of Markov chain Monte Carlo (MCMC) samplings in order to infer the posterior probability distribution of cosmological, baryonic, intrinsic alignment, and two-body DDM parameters based on the WL data from {\tt KiDS} and the CMB observations from {\tt Planck}. We employed the stretch-move ensemble method implemented within the \texttt{emcee} package \citep{emcee} to sample from the posterior distribution. 

An overview of the sampled parameters and the prior choices are listed in Tab.~\ref{tab:mcmc_parameters}. We used flat priors for all cosmological parameters except for the optical depth $\tau$, where we assumed a Gaussian prior $\mathcal{N}(0.0506,0.0086)$, as explained in Section~3.1 of \cite{bucko_2022_1bddm}. For the dark matter abundance ($\omega_{\rm dm}$) and the amplitude of the primordial power spectrum ($A_{\rm s}$), we used priors that are wide enough to comfortably include the WL posteriors found by \cite{kids_1000}. We note that $\omega_{\rm dm}$ represents the initial dark matter abundance, and as we assume only late-time decays, this parameter describes  the total initial dark matter budget in both the $\Lambda$CDM and $\Lambda$DDM scenarios.  For the baryon abundance $\omega_{\rm b}$, Hubble constant $h_0$, and spectral index $n_{\rm s}$, which cannot be well constrained by WL alone, we choose the prior that is as wide as possible to span the values found by surveys (e.g. CMB data) that are more sensitive to these parameters. The prior range for the intrinsic alignment parameter $A_{\rm IA}$ is wide enough to include the posterior distribution of this parameter found by \cite{kids_1000}. The {\tt Planck} absolute calibration $A_{\rm planck}$ was probed under the Gaussian prior $\mathcal{N}(1.0,0.0025)$.\footnote{\url{https://wiki.cosmos.esa.int/planck-legacy-archive/index.php/CMB_spectrum_\%26_Likelihood_Code}} We further imposed flat priors on the baryonic parameters $\log_{\rm 10} M_{\rm c},\theta_{\rm ej}$, and $\eta_\delta$ covering the full range of the {\tt BCemu} parameters.
For the $\Lambda$DDM decay rate ($\Gamma$) and the velocity kick magnitude ($v_k$), we assumed flat priors for  both the $\log_{10} \Gamma$ and $\log_{10} v_k$ spanning from $\Lambda$CDM values up to the upper boundary limited by the range of the $\Lambda$DDM emulator (see Tab.~\ref{tab:mcmc_parameters}).

\begin{table}
    \tiny
    \renewcommand{\arraystretch}{1.3}
    \centering
    \begin{tabular}{l l l c}
    \hline
    \hline
    \rule{0em}{-0.5em}\\
    Parameter name & Acronym     & prior  & range  \\
    \hline
(Initial) DM abundance & $\omega_{\rm dm}$     	&	flat	&	[0.051, 0.255]\\ 
Baryon abundance & $\omega_{\rm b}$     	&	flat	&	[0.019, 0.026]\\ 
Scalar amplitude & $\ln (10^{10} A_{\rm s})$	&		flat	&	[1.0, 5.0]\\ 
Hubble constant &$h_0$	&	flat	&	[0.6, 0.8]\\ 
Spectral index &$n_{\rm s}$     &	flat	&	[0.9, 1.03]\\  
Optical depth &$\tau_{\rm reio}$	&	normal	&	$\mathcal{N}(0.0506,0.0086)$\\ 
\hline
Intrinsic alignment amplitude &$A_{\rm IA}$     	&		flat	&	[0.0, 2.0]\\ 
Planck calibration parameter &$A_{\rm planck}$	&		normal	&	$\mathcal{N}(1.0,0.0025)$\\ 
\hline
First gas parameter (\texttt{BCemu}) &$\log_{10} M_{\rm c}$     	&		flat	&	[11.0, 15.0]\\ 
Second gas parameter (\texttt{BCemu}) &$\theta_{\rm ej}$     	&		flat	&	[2.0, 8.0]\\ 
Stellar parameter (\texttt{BCemu}) &$\eta_\delta$     	&	 	flat	&	[0.05, 0.40]\\ 
\hline
Decay rate (Gyr$^{-1}$) & $\log_{10} \Gamma$	&		flat	&	[$-4.00$, $-1.13$]\\  
Velocity kick magnitude (km/s) & $\log_{10} v_k$ 	&		flat	&	[0.00, 3.69]\\ 
    \hline
    \end{tabular}
    \caption{Parameter samples in MCMC chains. Next to the parameter, we show the prior type used and the range for a flat prior or mean and standard deviation when a Gaussian prior was used. In the top part \jozef{of the table}, cosmological parameters are stated followed by the intrinsic alignment and Planck absolute calibration parameter, three baryonic parameters, and two free parameters characterising the decaying dark matter.}
    \label{tab:mcmc_parameters}
\end{table}

\begin{figure*}
    \centering
    \includegraphics[width = 0.85\textwidth]{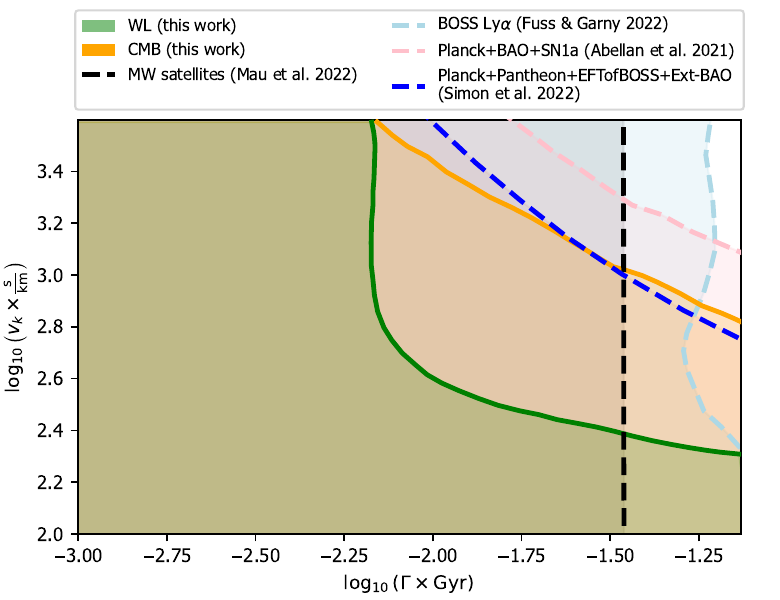}
    \caption{Two-dimensional constraints on $\Lambda$DDM parameters $\Gamma$ and $v_k$ as obtained from WL cosmic shear (green) and CMB (orange) data. We also display the results from several recent studies constraining $\Lambda$DDM parameters using different types of observables (see Section~\ref{sec:constraints} for more details).}
    \label{fig:Gamma_vk}
\end{figure*}

In both the WL and CMB setups, we assumed Gaussian likelihoods. In the case of the CMB, we used a marginalised light-weight version of the full Planck likelihood called  
{\tt Plik\_lite} \citep{Prince_2019_Plik_lite,planck_V_power_spectra_likelihoods}, as it is an affordable approximation in the case of $\Lambda$DDM. \cite{Abellan_2021} demonstrated that using {\tt Plik\_lite} for $\Lambda$DDM produces only a negligible difference on the recovered posteriors compared to a full {\tt Planck} analysis. Also, {\tt Plik\_lite} comes with the marginalised version of the covariance matrix, which we used throughout this work. In the case of WL, we used the band power covariance matrix as published in \cite{kids_1000}. To assess the convergence of our chains, we applied the \textit{Gelman-Rubin} criterion \citep{Gelman_Rubin_1992}, assuming the chains to be converged at $R_c<1.1$.

We used two different metrics to assess the tension between the CMB and WL observations within the assumed cosmological model. The first one is the standard Gaussian metric \citep[see e.g.][]{kids_1000}, which is defined as 
\begin{eqnarray}
    \label{eq:S8-tension-def}
    & &\tau_{S_8} = \frac{S_8^{\rm CMB} - S_8^{\rm WL}}{\sqrt{\rm{Var}\left[S_8^{\rm CMB}\right] + \rm{Var}\left[S_8^{\rm WL}\right]}},
\end{eqnarray}
where $\rm Var\left[\right]$ stands for the variance of the $S_8$ measurements from the two datasets. We note that the Gaussian metric assumes a Gaussian posterior distribution of the parameter of interest and, importantly, is agnostic about how well the underlying data are being fitted. Due to these shortcomings, we also employed the difference in maximum a posteriori $Q_{\rm DMAP}$ criterion \citep{raveri_hu_qmap_2019}. This metric is defined as
\begin{equation}
    \label{eq:QMAP}
    Q_{\rm DMAP} = \chi^2_{\rm min,WL+CMB} - \left(\chi^2_{\rm min,WL} + \chi^2_{\rm min,CMB}\right),
\end{equation}
where $\chi^2_{\rm min,WL}$, $\chi^2_{\rm min,CMB}$, and $\chi^2_{\rm min,WL+CMB}$ correspond to the minimum chi-squared values from the WL, the CMB, and the combined analysis. The $Q_{\rm DMAP}$ criterion evaluates the incapability of a combined analysis to approach the goodness of fit of the individual analyses, that is, when either of the datasets is fitted separately. In terms of $\sigma$, the tension is quantified as $\sqrt{ Q_{\rm DMAP} }$.


Although the emulator presented in Section~\ref{sec:emul_2bDDM} accounts for the three parameters $\Gamma$, $v_k$, and $f$, we fixed the fraction of decaying dark matter to stable dark matter to unity ($f=1$). This means that we restricted our analysis to the case of a universe with one initial (unstable) dark matter fluid, leaving the investigation of a multi-fluid dark matter sector to future work.

\section{Results}
\label{sec:results}

In the first part of this section, we describe the constraints on two-body decays from WL and CMB observations and report our findings regarding baryonic physics. In the second part, we revisit the impact of the $\Lambda$DDM model on the $S_8$~tension before concluding with a discussion on how well the observational data can be fitted within the $\Lambda$DDM model.

\subsection{Derived constraints on two-body decays}
\label{sec:constraints}

We obtained the constraints on the decay rate $\Gamma$ and velocity kick magnitude $v_k$ by marginalising over cosmology, baryons, and nuisance parameters. We display the outcome in Fig.~\ref{fig:Gamma_vk}, showing the $\Lambda$DDM posteriors at the 95\% credible intervals as obtained from the WL (green) and CMB  (orange) analyses. In both cases, the obtained constraints are in agreement with $\Lambda$CDM, showing no hint of decay in the dark matter sector. Figure~\ref{fig:Gamma_vk} also includes a comparison of our findings with several recent studies: \cite{Mau_2022} (black) focusing on Milky Way satellites, \cite{Fuss_Garny_2022} (light blue) examining the impact on Lyman-$\alpha$ forest, and \cite{Abellan_2021} and \cite{Simon_2022} (pink and blue, respectively) using {\it Planck 2018}, SNIa, and BAO data. We note that our CMB posteriors are consistent with previous CMB studies, in particular with the results of \cite{Simon_2022}. Our limits are slightly stronger than those of \cite{Abellan_2021} but exhibit similar hyperbolic contour trends. 
Milky Way satellites as probed by the DES collaboration \citep{Mau_2022} are sensitive to decay rate $\Gamma$ and rule out half-life times of $\tau<30$~Gyr, while the Lyman-$\alpha$ appears to be less sensitive to two-body decays than Milky Way satellites. Finally, the WL data alone provides the most stringent constraints, excluding regions where the following conditions hold at 95\% credible intervals:
\begin{eqnarray}
\tau = \Gamma^{-1} \lesssim 125~ \rm Gyr, and\\
v_k \gtrsim 300~ \rm km/s.~~~~
\end{eqnarray}
Hence, within the parameter space of $\Lambda$DDM and the datasets analysed in this study, WL data impose much stronger limits compared to CMB observations.\footnote{We find that it is important to note that in our analysis, we do not include the CMB lensing effect.} This is mainly due to the fact that, unlike for the case of one-body decays \citep{bucko_2022_1bddm}, the two-body $\Lambda$DDM model does not significantly affect the background evolution of the Universe, leaving the signal from the late-time integrated Sachs-Wolfe effect unchanged. More discussions about the effects of two-body decays on the CMB signal can be found in Appendix~\ref{app:parameter_effects}.

\begin{figure}[!ht]
    \centering
    \includegraphics[width = \columnwidth]{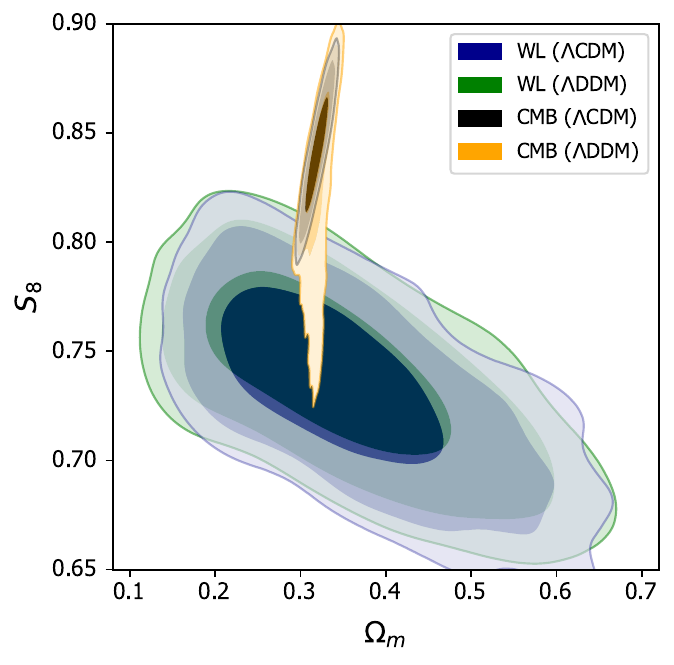}
    \caption{Two-dimensional posterior distributions of $\Omega_{\rm m}$ and $S_8$ as resulting from MCMC analysis highlighting the 68\%, 95\%, and 99\% confidence intervals. The WL results are shown in dark blue ($\Lambda$CDM) and  green ($\Lambda$DDM), while the findings based on CMB data are displayed as black ($\Lambda$CDM) and orange ($\Lambda$DDM) contours.}
    \label{fig:lcdm_vs_ddm_individual}
\end{figure}
\begin{figure}[!ht]
     \centering
    \includegraphics[width = \columnwidth]{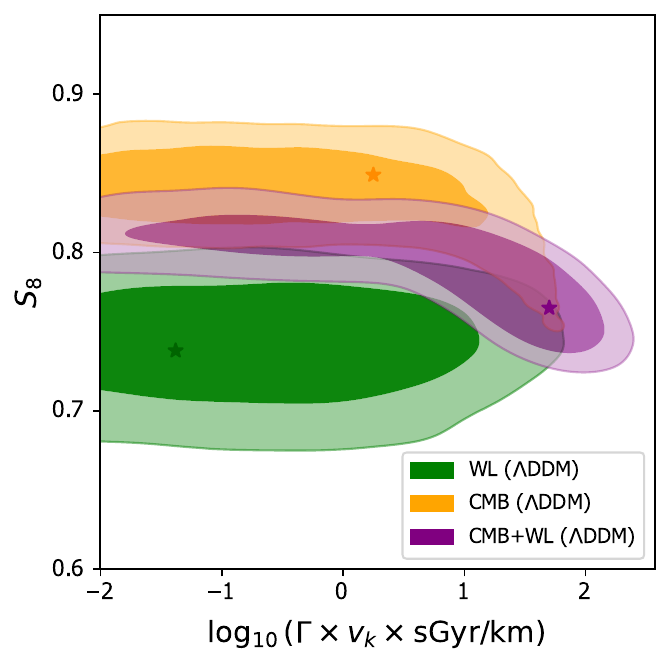}
        \caption{Effects of two-body decays on $S_8$ in the case of the WL (green contours), CMB (orange contours), and combined WL plus CMB data (purple contours). We show the 2D posterior contours of $\Omega_{\rm m}$ against the product of the DDM parameters at the 68\% and 95\% confidence level. We also display the best-fit model in each scenario as coloured stars.}
        \label{fig:Gamma_vk_S8}
\end{figure}

\subsection{Revisiting the $S_8$~tension between WL and the CMB}
As a next step, we investigated the extent the $\Lambda$DDM model is able to alleviate the $S_8$ tension between lensing and CMB data. In contrast to previous findings \citep[e.g.][]{Abellan_2021,abellan_2022,Tanimura:2023bkh}, we argue that the $\Lambda$DDM model is unable to significantly reduce the systematic shift between the clustering signal predicted from the CMB and the {\it KiDS-1000} WL survey. 

In Fig.~\ref{fig:lcdm_vs_ddm_individual}, we illustrate the $\Omega_{\rm m}-S_8$~posterior contours as obtained from the individual WL and CMB data analyses for both the $\Lambda$CDM (dark blue and black) and the $\Lambda$DDM model (green and orange). We observed only a marginal change in the WL and CMB contours when going from the $\Lambda$CDM to the $\Lambda$DDM case; both are at the 68\% and 95\% confidence level. However, at the 99\% confidence level, the CMB contours reveal a prominent feature pointing towards low $S_8$-values in apparent agreement with the WL analysis. 

Applying the $Q_{\rm DMAP}$ criterion defined in Eq.~\eqref{eq:QMAP}, we obtained a mutual tension of $3.4\sigma$ between WL and CMB data for the case of $\Lambda$CDM. When including two-body decays, the tension is reduced to $1.9\sigma$. Using the Gaussian metric defined in Eq.~\eqref{eq:S8-tension-def}, we obtained an $S_8$ tension of $3.0\sigma$ for the $\Lambda$CDM and $2.7\sigma$ for the $\Lambda$DDM. The difference between the two measures confirms the non-Gaussian shape of the CMB posteriors.

The origin of the non-Gaussian tail in the $\Omega_{\rm}-S_8$ plane (visible in Fig.~\ref{fig:lcdm_vs_ddm_individual}) can be better understood by looking at Fig.~\ref{fig:Gamma_vk_S8}, in which we plot the posterior distribution of $S_8$ against the product of the $\Lambda$DDM parameters ($\Gamma\times v_k$). One can see that for large values of $\Gamma\times v_k$, the posterior obtained from the CMB analysis bends downwards, preferring smaller values of $S_8$. However, this downturn happens at a part of the parameter space that is not favoured by the WL analysis. Instead of fully overlapping with the WL posteriors (green), the CMB contours (orange) rather wrap around them only, leading to a slight overlap of the 95\% confidence regions.

Based on the analysis above, we concluded that although the two-body $\Lambda$DDM alleviates the $S_8$ tension to some degree, it does not provide a convincing solution to it. Furthermore, the model does not yield improved fits to either the individual CMB or WL data (see Tab.~\ref{tab:chi2_values}) despite the two additional model parameters $\Gamma$ and $v_k$. For further tests of the $\Lambda$DDM model, we refer to Appendix~\ref{app:tension}.

Fitting the combined WL and CMB data, we obtained a preference for a model with non-zero DDM parameters. The best fitting values are $\log_{10} \Gamma = -2.25^{+0.74}_{-0.23}$ and $\log_{10} v_k > 2.80$, compatible with the overlap region of CMB and WL posteriors.
See Appendix~\ref{app:mcmc_results} for the full results from our MCMC analysis. These findings align with the conclusions of \cite{Simon_2022}, who added a Gaussian prior with the $S_8$ value from {\tt KiDS} to their CMB analysis. However, given the original tension between WL and CMB data, it is not surprising to obtain a preference for non-vanishing $\Lambda$DDM parameters in the combined data analysis. We want to stress that this is by no means a signal of a departure from $\Lambda$CDM but rather a natural consequence of the internal tension between the two datasets. This interpretation is confirmed by the fact that the posteriors from the individual analysis shown in Fig.~\ref{fig:lcdm_vs_ddm_individual} do not show significant overlap in the $\Lambda$CDM nor in the $\Lambda$DDM case.

The constraints inferred for all sampled parameters in our MCMC runs in the cases of CMB-only, WL-only, and the combined analysis can be found in Tab.~\ref{tab:lcdm_mcmc_results} and Tab.~\ref{tab:dcdm_mcmc_results} of Appendix~\ref{app:mcmc_results}.
In particular, the values of $S_8$ from our CMB and WL analyses are compared to the results of \cite{kids_1000}, \cite{schneider_2022_ksz} and \cite{planck_2018} in  Fig.~\ref{fig:S8_comparison}.  
Details about tests of our MCMC pipeline and the related discussion can be found in Appendix~A of \cite{bucko_2022_1bddm}.

\section{Summary and conclusion}
\label{sec:conclusion}
A dark matter scenario including particle decay forms a natural extension to the minimal model of a cold, stable, and collisionless dark matter particle. In this work, we investigated the case of two-body dark matter decays where particles decay into a massless particle and a massive daughter particle, the latter obtaining a velocity kick as a result of the decay process. This model has been studied in different contexts in the past \citep[e.g.][]{Peter_2bddm,Wang_Zentner_2012,Cheng_2015,Fuss_Garny_2022} and has been proposed as a potential solution to the $S_8$ tension \citep{Abellan_2021, abellan_2022, Simon_2022}.

In this paper, we performed the first fully non-linear analysis of WL and CMB data for the two-body decaying dark matter ($\Lambda$DDM) scenario. Based on a suite of $N$-body simulations, we constructed a neural network-based emulator to obtain fast predictions of non-linear matter power spectra for arbitrary values of the decay rate ($\Gamma$), the decay-induced velocity kick ($v_k$), and the fraction of decaying to total dark matter ($f$). We then included the emulator in our pipeline predicting WL observations and performed an MCMC analysis with WL data from {\it KiDS-1000} and CMB data from {\it Planck 2018}.

We present improved constraints on the two-body decaying dark matter parameters based on the WL data from {\it KiDS-1000}. Our constraints are significantly stronger compared to previous results. Specifically, we exclude models with $\tau = \Gamma^{-1} \lesssim 125$~Gyr and $v_k \gtrsim 300$~km/s. Figure~\ref{fig:Gamma_vk} provides a summary of our constraints from the WL and the CMB, along with previous results from the literature.

When considering the clustering (or $S_8$) tension between {\it KiDS-1000} and {\it Planck 2018}, we observed an improvement of \referee{1.5}$\sigma$ with $\Lambda$DDM as compared to the original \referee{3.4}$\sigma$ tension measured in the $\Lambda$CDM model\referee{ using $Q_{\rm DMAP}$ criterion, which can account for the non-Gaussianity of the posteriors compared to Gaussian tension better. However, the WL and CMB posteriors overlap at a very small confidence, and therefore, we} conclude that the two-body $\Lambda$DDM scenario is unable to convincingly resolve the clustering tension between WL and CMB observations. We note that previous works obtaining different conclusions \citep[e.g.][]{Abellan_2021,abellan_2022,Simon_2022,Tanimura:2023bkh} did not include a full, self-consistent modelling of the WL signal and were therefore unable to directly test the $S_8$ tension in the case of $\Lambda$DDM.

A further step forward with respect to the current analysis could involve the analysis of the $\Lambda$DDM model using DES observations along with the possible addition of galaxy clustering and galaxy-galaxy lensing. Furthermore, including additional low-redshift datasets, such as eBOSS and SNIa, could lead to stronger constraints on parameters such as $h_0$ and $\Omega_{\rm b}$, allowing for a more precise determination of the dark matter parameters. Finally, we expect data from Euclid and the Vera C. Rubin Observatory to significantly improve current limits on dark matter decays.

The emulator of two-body decays developed in this study is now publicly available as the {\tt DMemu} Python package. We welcome researchers to incorporate this package into their data analysis pipelines and further test the $\Lambda$DDM model.

\begin{figure}
    \centering
    \includegraphics[width = \columnwidth]{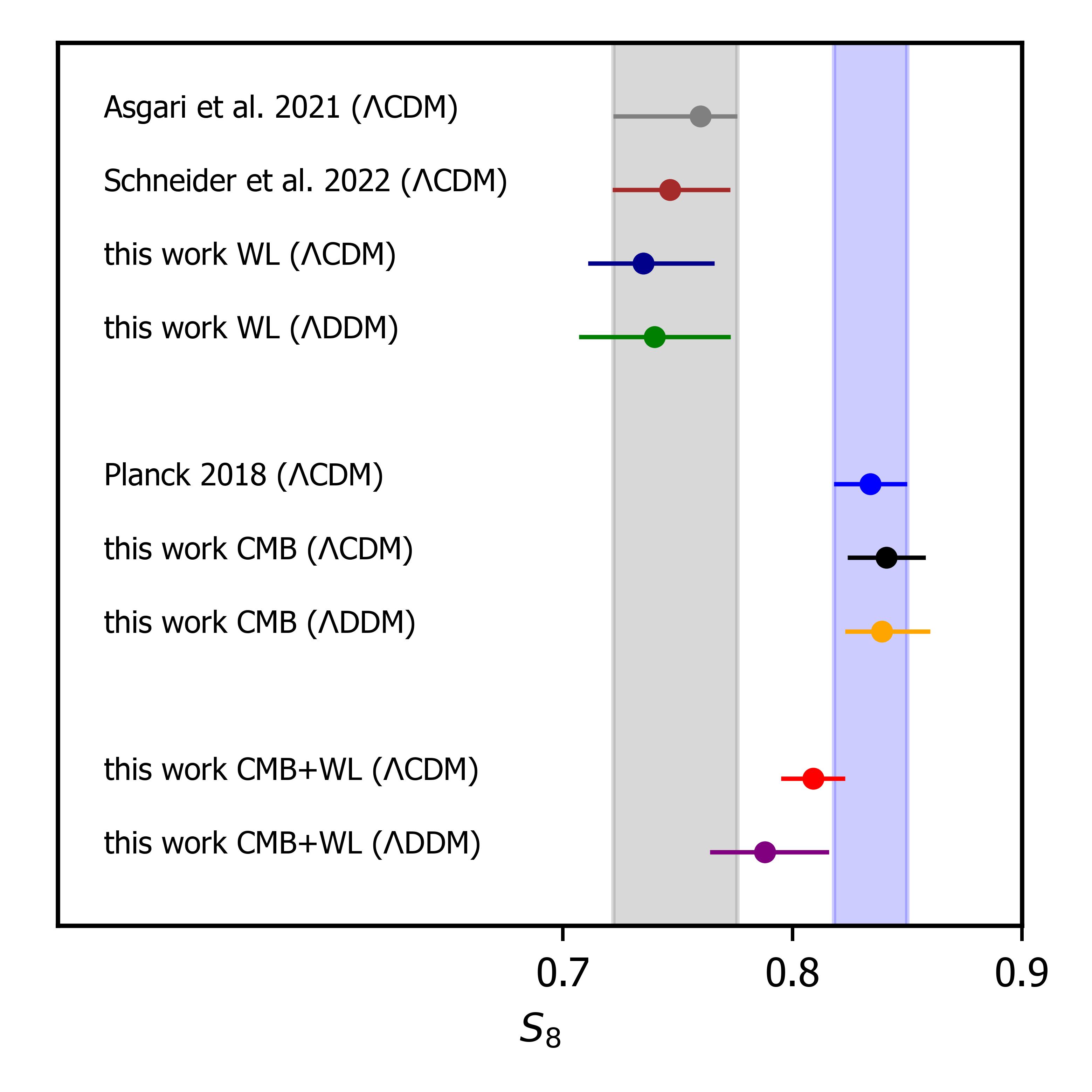}
    \caption{One-dimensional constraints of the $S_8$ parameter as inferred from the CMB, WL, and combined analyses. The original results from the {\it KiDS-1000} \citep{kids_1000} and the {\it Planck 2018} \citep{planck_2018} analyses are included in grey and blue for comparison. We also show the results of \cite{schneider_2022_ksz} using a similar WL modelling recipe.}
    \label{fig:S8_comparison}
\end{figure}

\begin{acknowledgements}
We thank Douglas Potter and Jonathan Hubert for helpful discussions and technical support. This work is supported by the Swiss National Science Foundation under grant number PCEFP2\_181157. 
Nordita is supported in part by NordForsk.
\end{acknowledgements}

%
%
\bibliography{2bDDM.bib}
\bibliographystyle{aa}

\appendix

\section{Comparing two-body decay simulations to previous work}
\label{app:compare_model}

\begin{figure}
    \centering
    \includegraphics{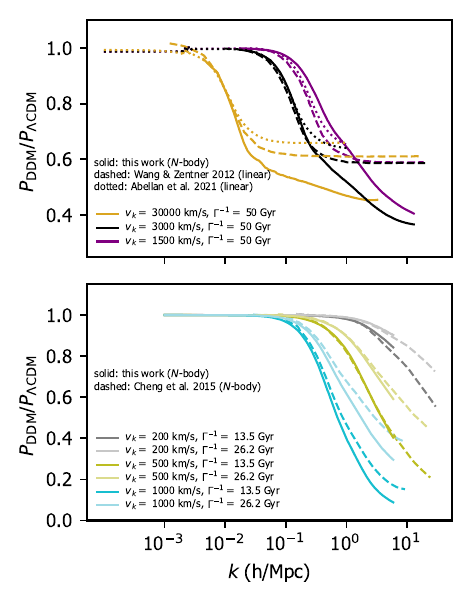}
    \caption{Comparison of our $N$-body simulations to previous works studying the $\Lambda$DDM model. The top panel shows how our simulations (solid lines) compare to linear (Boltzmann) codes of \cite{Wang_Zentner_2012} (dashed lines) and \cite{Abellan_2021} (dotted lines) for three different decaying dark matter configurations. The bottom panel shows simulations of two-body decays published in \cite{Cheng_2015} compared to our results for different decay rates $\Gamma$ and velocity kick magnitudes $v_k$.}
    \label{fig:sims_benchmark}
\end{figure}

In Fig.~\ref{fig:sims_benchmark}, we present a comparison of our $N$-body simulations with three other studies that have investigated the same $\Lambda$DDM model \citep{Wang_Zentner_2012,Cheng_2015,Abellan_2021}. In the top panel of Fig.\ref{fig:sims_benchmark}, we compare linear recipes based on solving the Boltzmann hierarchy, as developed by \cite{Wang_Zentner_2012} (dashed lines) and \cite{Abellan_2021} (dotted lines). The solid lines represent the results of our $\Lambda$DDM $N$-body simulations for $f=1.0$.
For the chosen values of $v_k$ and $\Gamma$, we observed good agreement in terms of the downturn scales. However, at non-linear scales, there are more pronounced differences between the results, which is expected due to the limitations of linear calculations at higher values of $k$. We note that for $v_k=30000$ km/s, which corresponds to 10\% of the speed of light, we have concerns about the accuracy of our $N$-body implementation. Therefore, we did not perform model inferences for such extreme values of $v_k$ in this work.

In the bottom panel of Fig.~\ref{fig:sims_benchmark}, we provide a benchmark of the $\Lambda$DDM model (for $f=1.0$) by comparing it with the N-body study conducted by \citet{Cheng_2015} for six different sets of dark matter parameters. We found that scenarios with smaller velocity kicks ($v_k\leq 500$ km/s) exhibit agreement at the percent level across all scales.
At the non-linear regime ($k \gtrsim 1$ h/Mpc) and for larger velocity kicks ($v_k\geq 1000$ km/s), we observed larger deviations in the predicted power suppression. These deviations can be attributed to the different dark matter implementations employed by \citet{Cheng_2015}. However, it is important to note that scenarios with such a significant power suppression in the non-linear regime are ruled out by observations, which favour a much weaker decrease of power at $k\approx 1 - 10$ h/Mpc \citep{Wang_Zentner_2012}. Therefore, the observed differences between our results and those of \citet{Cheng_2015} are not a cause for concern.

\section{Cosmology dependence of suppression of the matter power spectrum in $\Lambda$DDM with respect to $\Lambda$CDM}
\label{app:cosmology_dependence}

The effects of $\Lambda$DDM in our study are accounted for by applying a cosmology-independent boost to the $\Lambda$CDM non-linear matter power spectrum, calculated using the {\tt revised\_halofit} method within the {\tt CLASS} code. To demonstrate the validity of this approach, we conducted a test suite of $N$-body simulations where we kept the $\Lambda$DDM parameters fixed at $\Gamma^{-1} = 26.20$ Gyr and $v_k = 500$ km/s and varied one $\Lambda$CDM parameter at a time. 

\begin{figure*}
    \includegraphics[width = \textwidth]{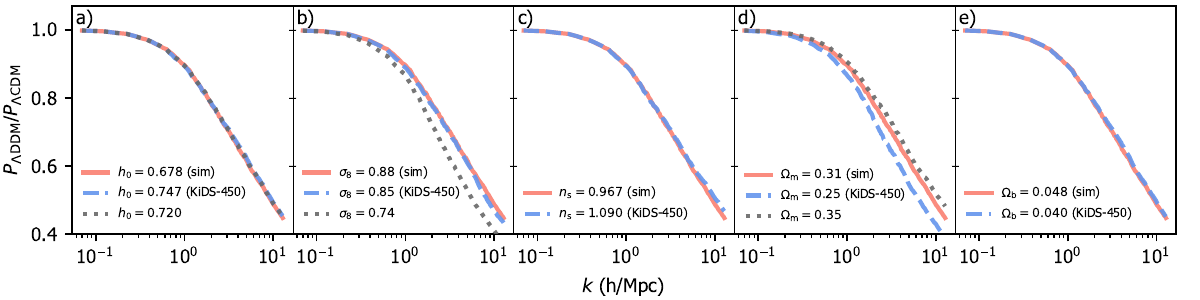}
    \caption{Cosmology dependence of the two-body effects on the matter power spectrum ratio between the $\Lambda$DDM and $\Lambda$CDM models. From panels (a) to (e), respectively, we show the effects of Hubble parameters ($h_0$), clustering amplitude ($\sigma_8$), spectral index ($n_{\rm s}$), matter abundance ($\Omega_{\rm m}$), and baryon abundance ($\Omega_{\rm b}$). With a solid salmon line, we show the fiducial cosmology of our $N$-body simulations, while the cosmology of the {\tt KiDS-450} survey is shown with a dashed blue line. We have also added a few additional scenarios, represented by dotted grey lines.}
    \label{fig:two-body-cosmo-dep}
\end{figure*}

We show in Fig.~\ref{fig:two-body-cosmo-dep} that this approach provides a good approximation, as neglecting the cosmology dependence in the applied boost results only in a second-order effect. In panels (a) to (e) of this figure, we show the impact of the Hubble parameter ($h_0$), clustering amplitude ($\sigma_8$), spectral index ($n_{\rm s}$), matter abundance ($\Omega_{\rm m}$), and baryon abundance ($\Omega_{\rm b}$), respectively. We display the $\Lambda$DDM boost for the fiducial cosmology of our $N$-body simulations with a solid salmon line, while we plot the same quantity for the best-fit cosmology of the {\tt KiDS-450} survey \citep{hildebrandt_2016_kids_450} with a dashed blue line. We also added a few additional models, represented by the dotted grey lines. 

In this analysis, we found that the choice of cosmology has a noticeable impact on the $\Lambda$DDM boost only for the parameters $\sigma_8$ (panel b) and $\Omega_{\rm m}$ (panel d). For the remaining parameters, the choice of cosmology does not significantly affect the boost. We find it is important to note that the models exhibiting substantial differences assume cosmologies that are quite distinct from the fiducial one. However, even in the cases of $\sigma_8$ and $\Omega_{\rm m}$, the observed effects are relatively small compared to the overall amplitude of the observed boost.

\section{Effect of model and baryonic feedback parameters on the observables}
\label{app:parameter_effects}

\subsection{Weak lensing}
To correctly interpret our results, it is important to understand the role of model parameters in relation to observables. Figure~\ref{fig:cls_modif_from_2bddm} shows the original {\it KiDS-1000} measurements of cosmic shear band powers with corresponding error bars (black points), and  we illustrate how well the WL, CMB, and WL plus CMB scenarios can fit these data. The solid green lines show the best-fit configuration in the WL-only setup in the case of $\Lambda$DDM (however, it performs very similar to the $\Lambda$CDM case). Once we calculated the band powers resulting from the CMB-only best-fit cosmology (orange lines), we recovered a significantly stronger WL signal compared to the {\tt KiDS} observations. This is the consequence of larger clustering ($S_8$~value) being preferred by the CMB data. In the combined WL plus CMB scenario, tighter error bars on the cosmological parameters from {\tt Planck} result in a best-fit cosmology that is close to the one from the CMB-only scenario, but baryons and, namely, two-body decays on small scales provide a much better fit to the cosmic shear data compared to the CMB-only scenario, as seen from the solid purple lines in the figure. To also showcase that two-body decays have a significant influence on the shear signal, we display the dashed and dotted purple lines. The former represents the same cosmology and baryonic parameters as in the solid purple case but switching off the two-body decays completely, while the latter displays the largest possible impact of the two-body decays in the case of cosmology and baryons fixed to the WL plus CMB $\Lambda$DDM best-fit scenario (solid purple lines). This impact is large enough to completely over- or underestimate the measured shear signal, as can be seen from the autocorrelation spectra of higher-redshift bins, for example.  This also explains why in the case of the combined $\Lambda$DDM analysis, we recovered the constraints on the two-body parameters that are detached from $\Lambda$CDM cosmology.


\subsection{Baryon feedback}

\begin{figure}
    \centering
    \includegraphics[width = 1\columnwidth]{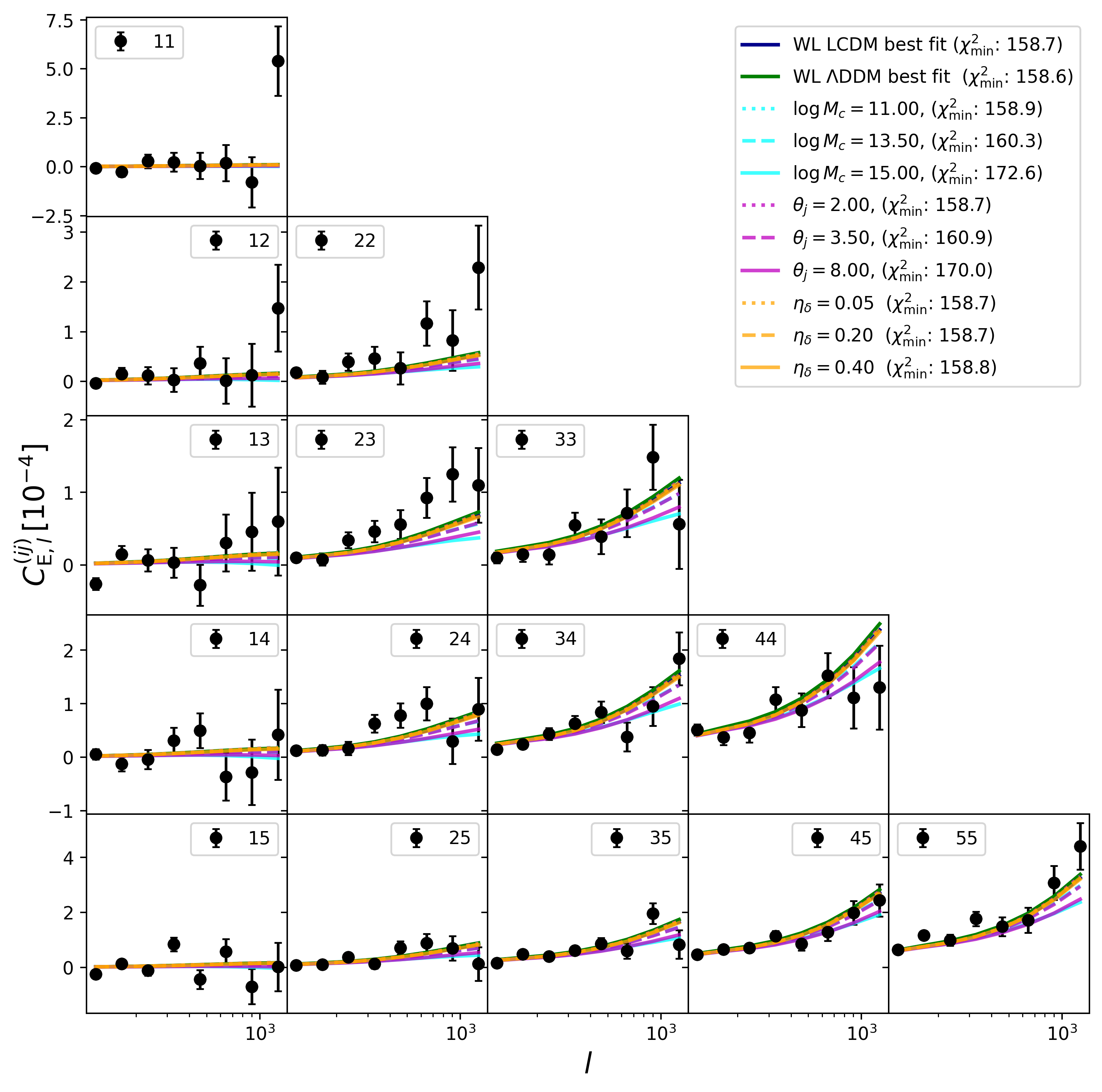}
    \caption{Effects of  baryons on band power spectra illustrated using different baryonic parameter configurations. Band power auto- and cross-correlations for five tomographic bins with respective error bars as measured by the {\it KiDS-1000} survey are shown as black dots. Solid dark blue (green) lines represent the best fit in the case of $\Lambda$CDM ($\Lambda$DDM), and the cyan, magenta, and orange lines depict the model predictions for different modifications of the $\Lambda$CDM best fit, varying baryonic parameters. In the legend, we include the minimum $\chi^2$ value for every plotted configuration.}
    \label{fig:cls_modif_from_baryons}
\end{figure}

We used three free parameters in the baryonic effects  emulator ($\log_{10} M_{c\rm },\theta_{\rm ej},\eta_\delta$) as proposed in \cite{Giri_2021_bcemu}. From our analysis, we found that the WL data does not provide any information about the stellar population parameter $\eta_\delta$. This is evident from Fig.~\ref{fig:cls_modif_from_baryons} and can be inferred from Tab.~\ref{tab:dcdm_mcmc_results}, as varying $\eta_\delta$ does not affect the modelled WL observables.
On the other hand, the gas profile parameters $\log_{10} M_{\rm c}$ and $\theta_{\rm ej}$ have an impact on the WL observables. For $\log_{10} M_{\rm c}$, we obtained $\log_{10} M_{\rm c}<13.1, (13.2)$ for $\Lambda$CDM ($\Lambda$DDM) from the WL-only analysis and $\log_{10} M_{\rm c}>13.8, (unconst)$ from the combined analysis. Similarly, for $\theta_{\rm ej}$, we found $\theta_{\rm ej} < 5.45, (5.57)$ in the case of $\Lambda$CDM ($\Lambda$DDM) from the WL-only analysis and $\theta_{\rm ej} > 5.88, (unconst)$ from the combined analysis. These results indicate that the values of these baryonic parameters are mutually exclusive when comparing the WL-only and combined ($\Lambda$CDM) scenarios. The underlying reasons for this behaviour are discussed in the following sections.

Fig.~\ref{fig:cls_modif_from_baryons} demonstrates the impact of baryonic feedback on the {\tt KiDS} band powers by varying the $\log M_c, \theta_{\rm ej}$, and $\eta_\delta$ parameters within the {\tt BCemu} framework. The best-fit configurations for $\Lambda$CDM ($\Lambda$DDM) are depicted in dark blue (green), while the coloured lines represent the variation of baryonic feedback strength in the $\Lambda$CDM best-fit case. Notably, we observe that modifying the stellar population parameter $\eta_\delta$ does not alter the shear signal. However, adjustments to the gas profile parameters $\log_{10} M_c$ and $\theta_{\rm ej}$ affect the model predictions at scales $l>300$.

\begin{figure}[!t]
    \centering
    \includegraphics[width = 1\columnwidth]{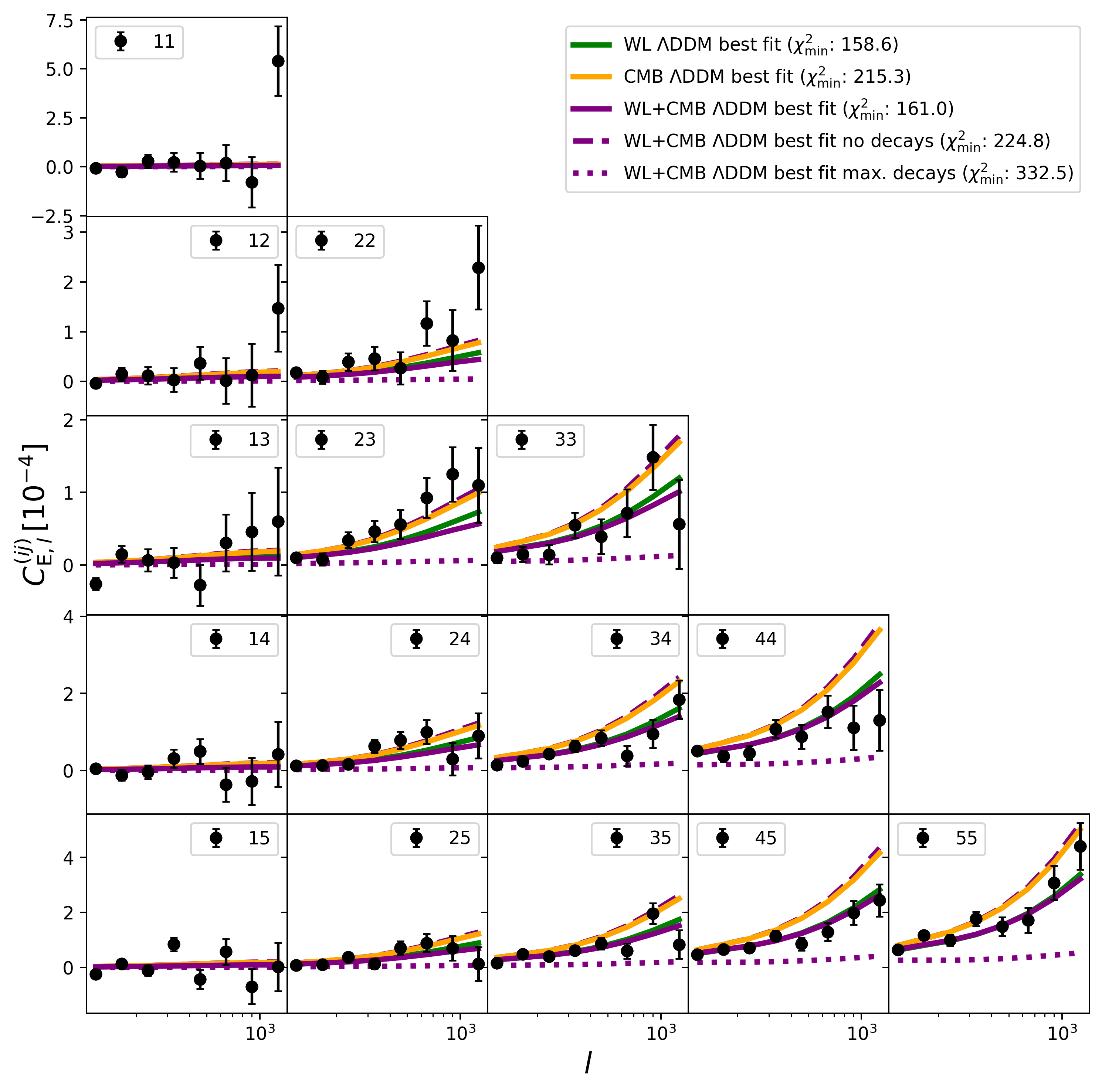}
    \caption{Effects of two-body decays on band power spectra illustrated using different forward-modelled $\Lambda$DDM configurations. Band power auto- and cross-correlations for five tomographic bins with respective error bars as measured by the {\it KiDS-1000} survey are shown as black dots. Solid green (orange)  lines represent the best-fit configurations of the WL (CMB) analysis projected on the band powers. The purple lines represent the different configurations of the WL plus CMB setup in $\Lambda$DDM cosmology. The solid lines show the best fit, while the dashed and dotted lines respectively illustrate the impact of underestimating and overestimating the two-body decays, keeping all the remaining parameters fixed to those of the WL plus CMB best fit. In the legend, we include the minimal $\chi^2$ value for every plotted configuration.}
    \label{fig:cls_modif_from_2bddm}
\end{figure}

Fig.~\ref{fig:cls_modif_from_2bddm} and~\ref{fig:cls_modif_from_baryons} provide insights into how the baryonic and $\Lambda$DDM parameters impact the band powers. Both sets of parameters exhibit a similar qualitative effect, resulting in a suppression of the signal at small scales. However, the influence of baryons is relatively subtle compared to the effects of DDM, as indicated by the available priors.
Also, we note that baryonic effects primarily manifest on the smallest scales, with no discernible impact below $l \sim 300$. In contrast, $\Lambda$DDM can influence the signal on larger scales, particularly when considering scenarios with large velocity kicks.

The analysis thus reveals that when considering {\tt KiDS} data alone, the preferred values for baryonic feedback parameters (and in the $\Lambda$DDM case, the two-body DDM parameters) tend towards the lower end. However, when combined with CMB data, these parameters drive the cosmology towards higher $S_8$ values, resulting in an excessive boost to the WL signal, as discussed earlier. The baryonic parameters can partially counteract this effect in the $\Lambda$CDM scenario by decreasing the signal on small scales, which explains the extreme values observed for the gas profile parameters in the combined analysis with CMB. Nevertheless, the limited suppression capability of baryons is inadequate to fully accommodate the WL signal, leading to deteriorated fits for both WL and CMB.

\begin{figure}[!htb]
     \centering
     \begin{subfigure}[b]{0.45\columnwidth}
         \centering
         \includegraphics[width=\textwidth]{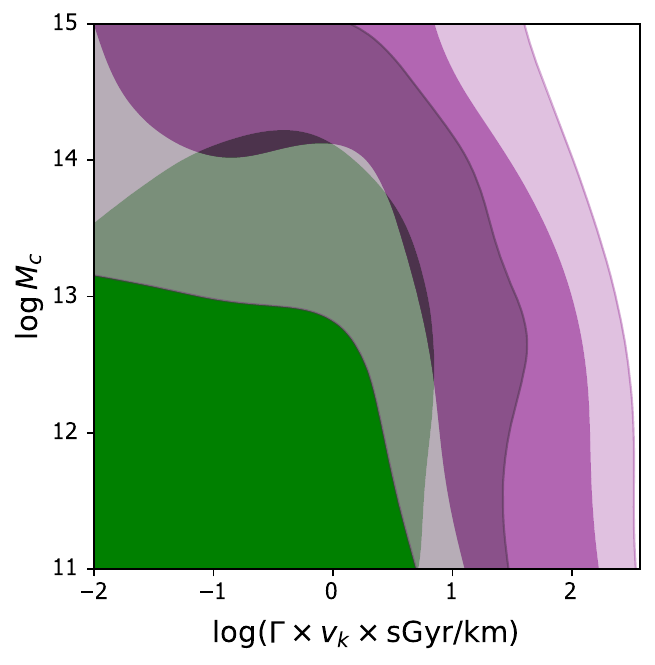}
         \label{fig:Gvk_logMc}
     \end{subfigure}
     \begin{subfigure}[b]{0.45\columnwidth{}}
         \centering
         \includegraphics[width=\textwidth]{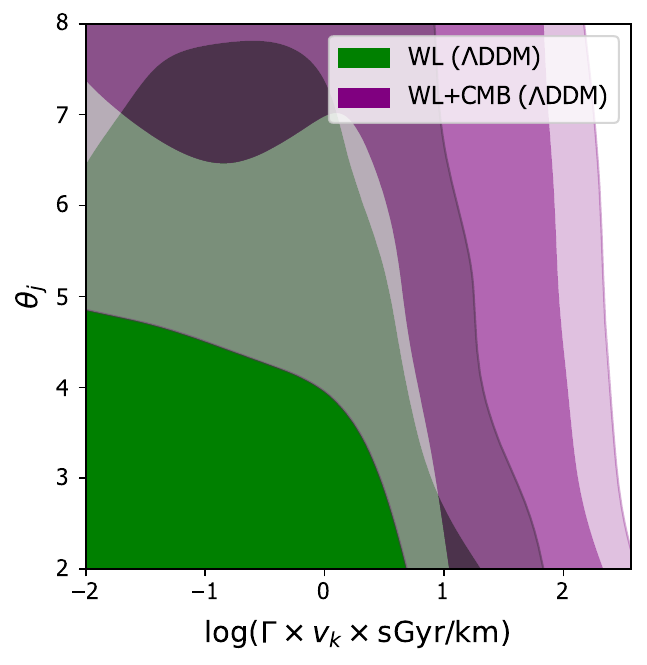}
         \label{fig:Gvk_thej}
     \end{subfigure}
        \caption{Effects of baryons and two-body decays in the $\Lambda$DDM universe as obtained from probing the WL (green) and WL plus CMB (purple) scenarios. Both baryons and two-body decays affect the matter power spectrum in qualitatively similar ways, and we observed a preference for weak suppression in the WL setup while requiring much stronger impact in the combined WL plus CMB analysis.}
        \label{fig:ddm_vs_baryons}
\end{figure}

In contrast, in the $\Lambda$DDM scenario, the DDM parameters can more effectively suppress the band power signal, resulting in improved fits for both WL and CMB compared to the $\Lambda$CDM setup. This can be observed in Fig.~\ref{fig:ddm_vs_baryons}, where the left panel depicts the combination of DDM parameters with $\log_{10} M_c$ and the right panel illustrates their combination with $\theta_{\rm ej}$ in the $\Lambda$DDM model. In both panels, it is evident that fitting the WL-only data favours a weak suppression of the power spectrum, indicated by the green contours located in the bottom-left region. Conversely, when combining WL and CMB, strong suppression is required either from baryons or from DDM, as indicated by the purple contours. It is conceivable that having even stronger baryonic effects at our disposal could entirely replace the need for the DDM and provide a satisfactory fit to both datasets individually, though the physical justification for such high baryonic feedback remains an open question.

\subsection{Cosmic microwave background}

\begin{figure*}
     \centering
     \begin{subfigure}[b]{0.3\textwidth}
         \centering
         \includegraphics[width=\textwidth]{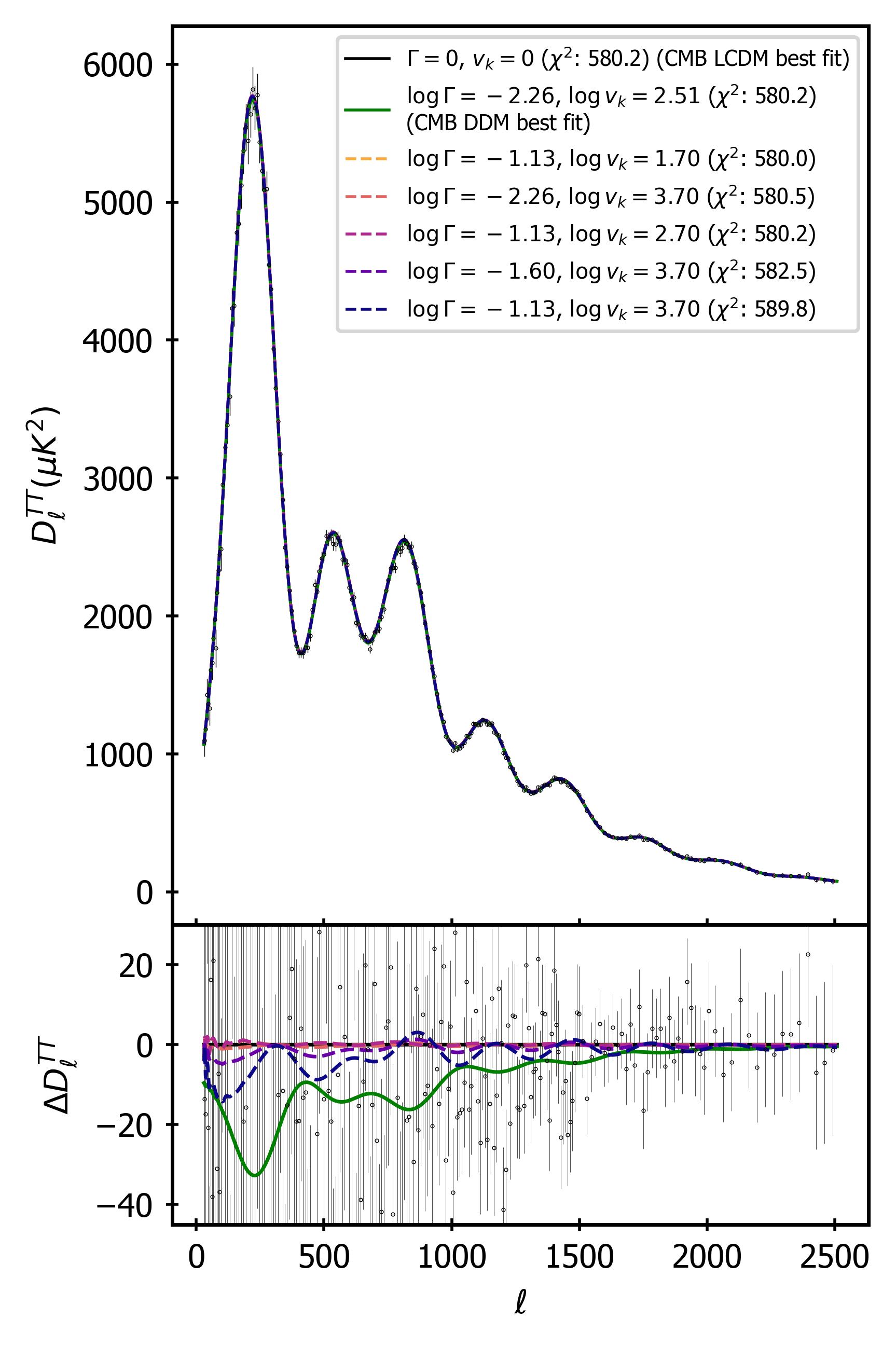}
         \label{fig:planck_tt}
     \end{subfigure}
     \begin{subfigure}[b]{0.3\textwidth}
         \centering
         \includegraphics[width=\textwidth]{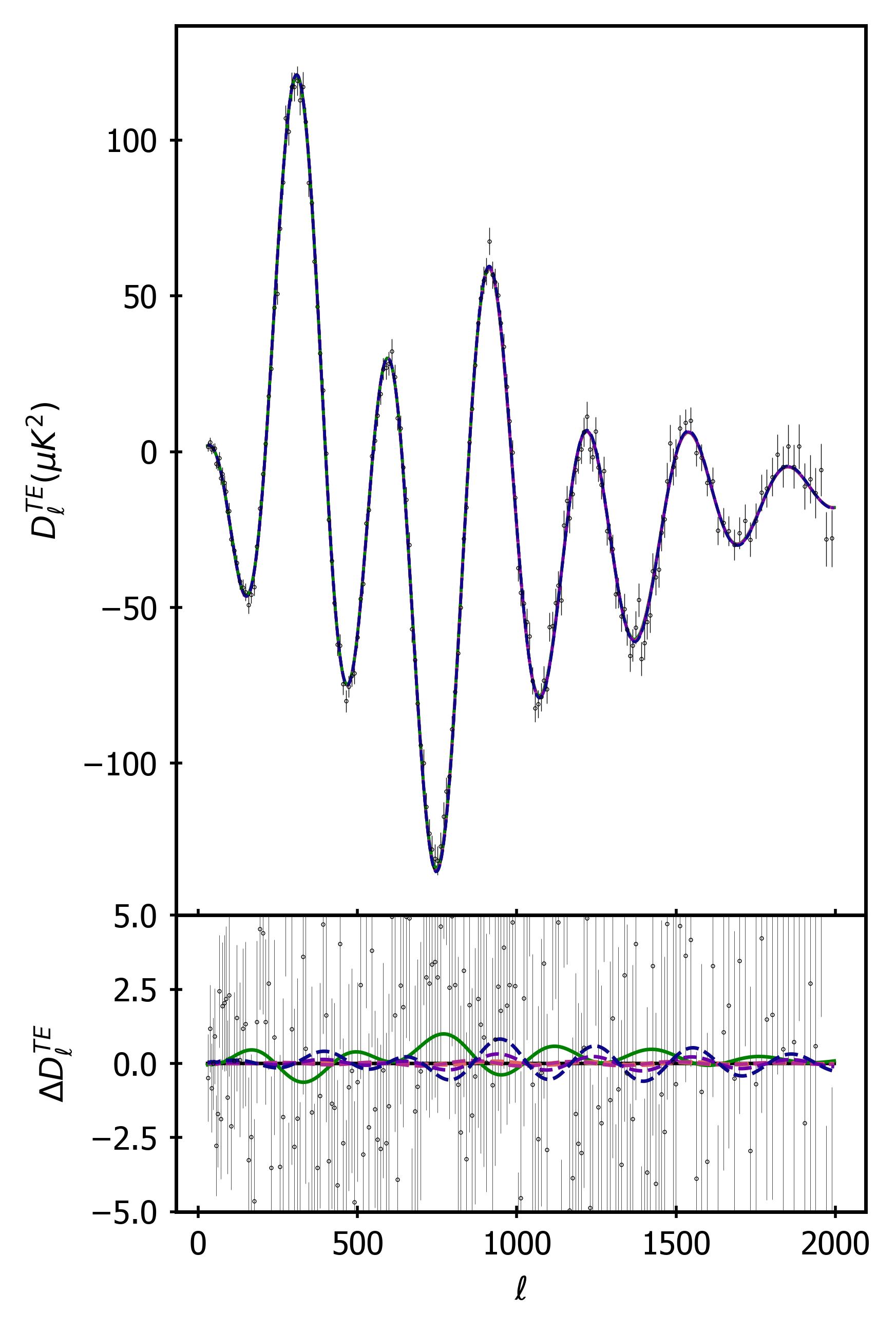}
         \label{fig:planck_te}
     \end{subfigure}
     \begin{subfigure}[b]{0.3\textwidth}
         \centering
         \includegraphics[width=\textwidth]{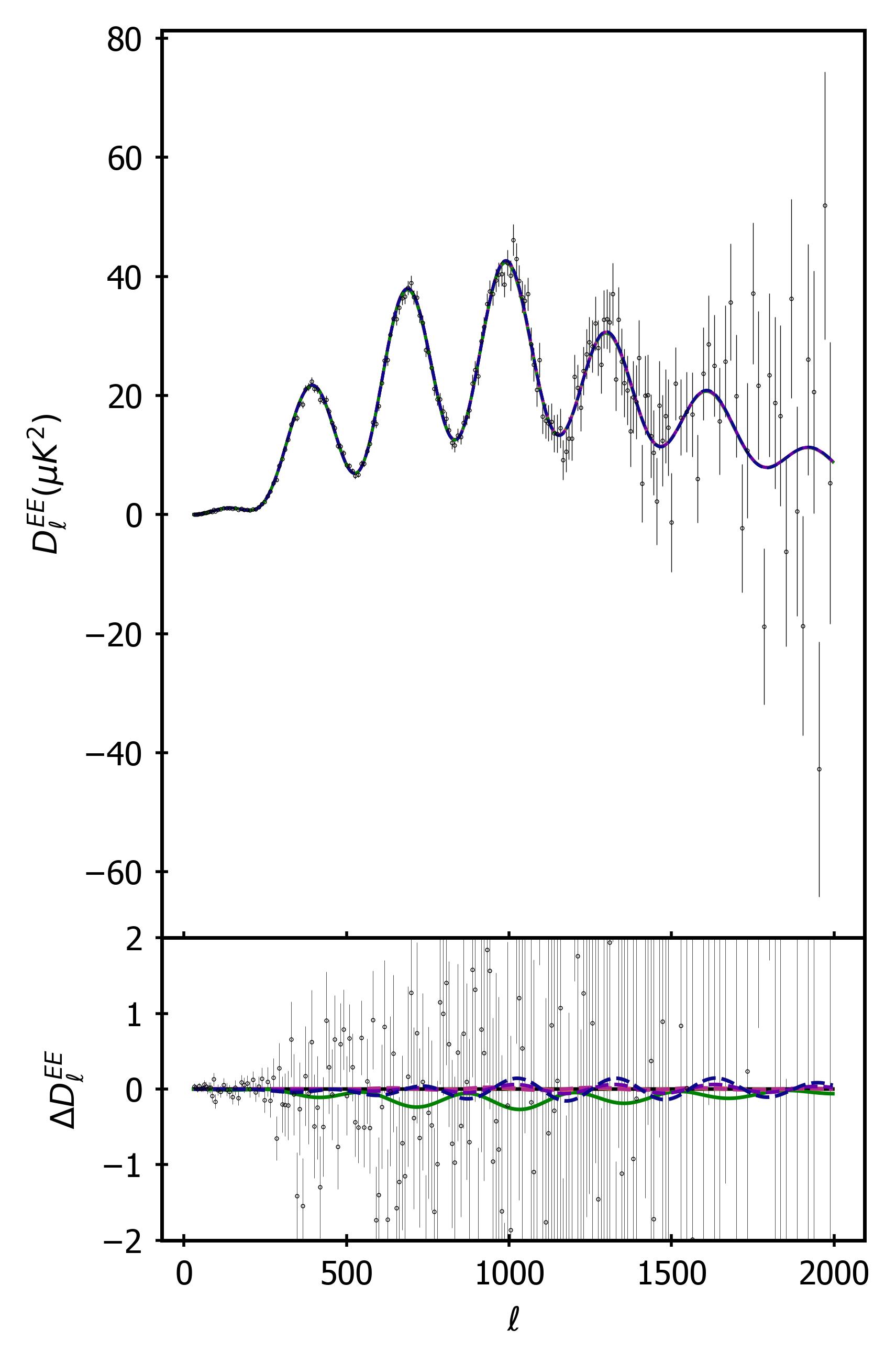}
         \label{fig:planck_ee}
     \end{subfigure}
        \caption{Planck TTTEEE angular power spectra as affected by different decaying dark matter parameter configurations. Black (green) solid lines represent the best fit in the case of $\Lambda$CDM ($\Lambda$DDM), and dashed curves depict the model predictions for
        different modifications of the $\Lambda$CDM best fit varying DDM parameters. In the legend, we include the $\chi^2$ value that each configuration yields.}
        \label{fig:planck_ddm_params}
\end{figure*}

In Fig.~\ref{fig:planck_ddm_params}, we show how two-body decays influence the predictions for CMB observables. We display temperature power spectra in the left panel and polarisation spectra in the right panel and add their cross-correlations in the middle panel. In the top row, we show the CMB spectra for $\Lambda$CDM (black) and $\Lambda$DDM (green) best fits, while we plot the predictions for the $\Lambda$CDM best-fit configuration extended by different combinations of $\Gamma$ and $v_k$ (dashed lines).  In the smaller bottom panels, we show the difference between all these predictions with the $\Lambda$CDM best-fit cosmology. We observed that the two-body decays do not change the CMB signal significantly, meaning that the scatter of the CMB data is larger than the observable effects of two-body decays. 


\section{Additional quantification of the tension}
\label{app:tension}
In the main text, we introduced two different metrics to estimate tension within a model at hand, namely, the Gaussian tension, defined in Eq.~\eqref{eq:S8-tension-def}, and the difference in maximum a posteriori, following Eq.~\eqref{eq:QMAP}. The latter criterion is not subject to the assumption of Gaussian posteriors but cannot deal with overfitting (i.e. the number of additional degrees of freedom our new model possesses). The last criterion we used to assess the efficiency of the $\Lambda$DDM model is the change in the Akaike information criterion (AIC) defined as 
\begin{eqnarray}
\label{eq:AIC}
& & \Delta {\rm AIC} = \Delta \chi^2_{\rm min} + 2(N_{\Lambda\rm DDM} - N_{\rm \Lambda CDM}),
\end{eqnarray}
where $N_\mathcal{M}$ is a number of free parameters in model $\mathcal{M}$. To determine whether a new model is preferred over the $\Lambda$CDM, substantial evidence against it is required based on {\it Jeffreys' scale} \citep{jeffreys_1961} using  $p<\exp{\left(-\Delta {\rm AIC}/2\right)}$ \citep{SCHONEBERG20221}, where $p = 10^{0.5}$. This implies that the difference in AIC ($\Delta {\rm AIC}$) between the new model and $\Lambda$CDM should satisfy the condition
\begin{equation}
\Delta {\rm AIC} < \Delta {\rm AIC}_0 = -2.3.
\label{eq:DAIC_crit}
\end{equation}

\begin{table}[]
    \renewcommand{\arraystretch}{1.3}
    \centering
    \begin{tabular}{l|c|c|c|c}
                        & WL   & CMB & Combined   & $\sqrt{ Q_{\rm DMAP}}$  \\
        \hline
        \hline
        $\chi^2_{\rm min}$ ($\Lambda$CDM)    & 158.7  &  580.2     & 750.5      & 3.4$\sigma$ \\ 
        $\chi^2_{\rm min}$ ($\Lambda$DDM)             & 158.6  &  580.2     & 742.6      & 1.9$\sigma$ \\ 
        $\Delta \chi^2_{\rm min}$ & -0.1    &  0.0       &  -7.9     & -- \\
        $\Delta {\rm AIC}$  & 3.9 & 4.0 & -3.9 & --\\
    \end{tabular}
    \caption{Minimum $\chi^2$ values resulting from sampling with KiDS and Planck data separately as well as from the combined analysis. In the last column, the tension between the two datasets is shown by the difference in maximum a posteriori criterion. The last two rows show differences of $\chi^2_{\rm min}$ between the $\Lambda$CDM and $\Lambda$DDM cases, subtracting the former from the latter, and the difference in Akaike information criterion $\Delta {\rm AIC}$ between the $\Lambda$CDM and $\Lambda$DDM scenarios. }
    \label{tab:chi2_values}
\end{table}
From the above equation, we obtained $\Delta {\rm AIC} = 3.9$ and $\Delta {\rm AIC} = 4.0$ in the WL and CMB scenarios, respectively, and this adding of two free parameters did not lead to an improved fit to the data. In the combined scenario,   $\Delta {\rm AIC} = -3.9< \Delta {\rm AIC}_0$. This means that the two additional parameters in the combined analysis are efficient. However, the combined $\Lambda$DDM fit is still worse by $\Delta \chi^2_{\rm min} = 3.8$ compared to what the $\Lambda$DDM scenario yields when treating the two datasets separately. We believe that fitting the data separately provides a better indicator of resolving the $S_8$~tension, in which the $\Lambda$DDM model fails. \referee{Throughout the work, we obtained the value of $\chi^2_{\rm min}$ as a minimum $\chi^2$ among all sampled configurations in our MCMC chain.} For a summary of combined MCMC analysis results, we refer to Tab.~\ref{tab:chi2_values}.

\section{MCMC results}
\label{app:mcmc_results}

In this section, we provide more details about the parameters resulting from our MCMC analysis. In Tab.~\ref{tab:lcdm_mcmc_results}, we report the findings from our $\Lambda$CDM analysis \citep[taken from][]{bucko_2022_1bddm}, while in Tab.~\ref{tab:dcdm_mcmc_results}, we summarise the actual $\Lambda$DDM model results. In both cases, we provide separate constraints from the WL, CMB, and combined analyses. In the upper part of the tables, we show the posteriors of the parameters sampled {by} \jozef{in our} MCMC \jozef{runs}, showing mean (best fit) values with corresponding upper and lower deviations. The dashed fields in the tables indicate the parameters were not present in a given MCMC setup, 'unconst' is used when parameter constraints could not be obtained from MCMC \jozef{analysis}, and we state no uncertainties whenever referring to a value constant throughout the inference.

\begin{table*}
    \tiny
    \renewcommand{\arraystretch}{1.3}
    \centering
    \begin{tabular}{l c c c}
    \hline
    \hline
    \rule{0em}{-0.5em}\\
         & WL $\rm \Lambda CDM$ & CMB $\rm  \Lambda CDM$ & WL + CMB $\rm \Lambda CDM$\\
    Parameter     & 68\% limits  & 68\% limits & 68\% limits \\
    \hline
$\omega_{\rm dm}$     	&	$	0.146 (0.081)^{+0.034}_{-0.056}	$	&	$	0.1208 (0.1200) \pm 0.0014	$	&	$	0.1182 ( 0.1188) \pm 0.0012	$	 \\ 
$\omega_{\rm b}$     	&	$	 \rm unconst (0.02455)	$	&	$	0.02231 (0.02236) \pm 0.00015	$	&	$	0.02248 (0.02253) \pm 0.00013	$	\\ 
$\ln (10^{10} A_{\rm s})$	&	$	 2.56 (3.50)^{+0.61}_{-0.80}	$	&	$	3.050 (3.110) \pm 0.017	$	&	$	3.039 (3.069)\pm 0.017	$	\\ 
$h_0$	&	$	 \rm unconst (0.6041)	$	&	$	0.6700 (0.6734) \pm 0.0061	$	&	$	0.6815 (0.6799)^{+0.0050}_{-0.0057}	$	\\
$n_{\rm s}$     	&	$	 \rm unconst (0.9377)	$	&	$	0.9622 (0.9640) \pm 0.0043	$	&	$	0.9678 (0.9668)\pm 0.0041	$	\\ 
$\tau_{\rm reio}$	&	$	-	$	&	$	0.0566 (0.0841) \pm 0.0083	$	&	$	0.0540 (0.0677) \pm 0.0077	$	\\ 
$A_{\rm IA}$     	&	$	0.75 (0.89)^{+0.33}_{-0.38}	$	&	$	- 	$	&	$	0.63 (0.44)^{+0.25}_{-0.31}	$	\\ 
$A_{\rm planck}$	&	$	-	$	&	$	1.0005 (1.0031) \pm 0.0025	$	&	$	1.0003 (1.0005) \pm 0.0025	$	\\ 
$\log_{10} M_{\rm c}$     	&	$	<13.1 (12.6)	$	&	$	-	$	&	$	 >13.8 (15.0)	$	\\ 
$\theta_{\rm ej}$     	&	$	<5.45 (2.23)	$	&	$	-	$	&	$	 >5.88 (7.74)	$	 \\ 
$\eta_\delta$     	&	$	 \rm unconst (0.21)	$	&	$	-	$	&	$	 \rm unconst (0.13)	$	\\ 
$\log_{10} (\Gamma \times \rm{Gyr})$	&	$	-	$	&	$	-	$	&	$	-	$	\\ 
$\log_{10} (v_k\times \rm{s/km})$	&	$	-	$	&	$	-	$	&	$	-	$	\\ 
\hline													
$\Omega_{\rm m}$     	&	$	0.347^{+0.066}_{-0.110}	$	&	$	0.3189 \pm 0.0087	$	&	$	0.3043 \pm 0.0070	$	\\ 
$\sigma_8$	&	$	 0.70^{+0.11}_{-0.13}	$	&	$	0.8154 \pm 0.0080	$	&	$	0.8029\pm 0.0073	$	\\ 
$S_8$	&	$	0.735^{+0.031}_{-0.024}	$	&	$	0.841 \pm 0.017	$	&	$	0.809 \pm 0.014	$	\\
\hline													
$\ln$(prior)     	&	$	5.99	$	&	$	7.9 (0.8)^{+1.2}_{-0.4}	$	&	$	13.9 (12.9)^{+1.0}_{-0.3}	$	\\ 
ln(${\rm lik}_{\rm WL}$)    	&	$	-81.3 (-79.3)^{+1.2}_{-0.5}	$	&	$	-	$	&	$	-83.5 (-83.3)^{+1.7}_{-1.0}	$	\\ 
ln(${\rm lik}_{\rm CMB}$)    	&	$	- 	$	&	$	-294.4 (-290.1)^{+2.0}_{-1.1}	$	&	$	-295.7 (-291.9)^{+2.8}_{-1.6}	$	\\ 
$\chi^2_{\rm min}$	&	$	158.7	$	&	$	580.2	$	&	$	750.5	$	\\ 
\hline													
    \end{tabular}
    \caption{Summary of our MCMC analysis assuming the $\Lambda$CDM model. We report individual results separately based on WL (\textit{KiDS-1000}) and CMB (\textit{Planck 2018}) data only as well as values inferred from the combined MCMC chain. We show the mean (best fit) values of the sampled (top) and derived (middle) parameters as well as the obtained prior, likelihood, and $\chi^2$~values (bottom).\protect \footnotemark }
    \label{tab:lcdm_mcmc_results}
\end{table*}

\footnotetext{adapted from \cite{bucko_2022_1bddm}}

\begin{table*}
    \tiny
    \renewcommand{\arraystretch}{1.3}
    \centering
    \begin{tabular}{l c c c}
    \hline
    \hline
    \rule{0em}{-0.5em}\\
         & WL $\Lambda\rm DDM$ & CMB $\Lambda\rm  DDM$ & WL + CMB $\Lambda\rm DDM$\\
    Parameter     & 68\% limits  & 68\% limits & 68\% limits \\
    \hline
$\omega_{\rm dm}$     	&	$	0.139 (0.106)^{+0.039}_{-0.055}	$	&	$	0.1208 (0.1190) \pm 0.0014	$	&	$	0.1191 (0.1193) \pm 0.0015	$	 \\ 
$\omega_{\rm b}$     	&	$	 \rm unconst (0.02103)	$	&	$	0.02231 (0.02242) \pm 0.00015	$	&	$	0.02241 (0.02245) \pm 0.00016	$	\\ 
$\ln (10^{10} A_{\rm s})$	&	$	 2.69 (2.93)^{+0.56}_{-0.94}	$	&	$	3.052 (3.116) \pm 0.017	$	&	$	3.045 (3.104)\pm 0.017	$	\\ 
$h_0$	&	$	 \rm unconst (0.6093)	$	&	$	0.6703 (0.6785) \pm 0.0062	$	&	$	0.6772 (0.6766) \pm 0.0067	$	\\
$n_{\rm s}$     	&	$	 \rm unconst (0.9295)	$	&	$	0.9620 (0.9667) \pm 0.0044	$	&	$	0.9653 (0.9665)\pm 0.0046	$	\\ 
$\tau$	&	$	-	$	&	$	0.0571 (0.0913) \pm 0.0082	$	&	$	0.0557 (0.0823) \pm 0.0085	$	\\ 
$A_{\rm IA}$     	&	$	0.77 ( 0.86) \pm 0.33	$	&	$	-	$	&	$	0.68 (0.90)^{+0.29}_{-0.33}	$	\\ 
$A_{\rm planck}$	&	$	-	$	&	$	1.0004 ( 1.0007) \pm 0.0025	$	&	$	1.0005 (1.0027) \pm 0.0024	$	\\ 
$\log_{10} M_{\rm c}$     	&	$	<13.2 (13.1)	$	&	$	-	$	&	$	 \rm unconst (12.4)	$	\\ 
$\theta_{\rm ej}$     	&	$	<5.57 (2.08)	$	&	$	-	$	&	$	 \rm unconst (2.67)	$	 \\ 
$\eta_\delta$     	&	$	 \rm unconst (0.11)	$	&	$	-	$	&	$	 <0.27 (0.27)	$	\\ 
$\log_{10} (\Gamma \times \rm{Gyr}) $	&	$	 \rm unconst (-2.12)	$	&	$	 \rm unconst (-2.26)	$	&	$	 -2.25 (-1.77)^{+0.74}_{-0.23}	$	\\ 
$\log_{10} (v_k\times \rm{s/km})$	&	$	 \rm unconst (0.74)	$	&	$	 \rm unconst (2.51)	$	&	$	>2.80 (3.47)	$	\\ 
\hline													
$\Omega_{\rm m}$     	&	$	0.338(0.341)^{+0.077}_{-0.11}	$	&	$	0.3200 (0.3084) \pm 0.0087	$	&	$	0.3100(0.3101)^{+0.0084}_{-0.0095}	$	\\ 
$\sigma_8$	&	$	 0.720(0.693)^{+0.094}_{-0.150}	$	&	$	0.812(0.837)^{+0.012}_{-0.0064}	$	&	$	0.776(0.752)^{+0.037}_{-0.031}	$	\\ 
$S_8$	&	$	0.740(0.738) \pm 0.033	$	&	$	0.839(0.849)^{+0.021}_{-0.016}	$	&	$	0.788(0.765)^{+0.028}_{-0.024}	$	\\
\hline													
ln(prior)     	&	$	3.6	$	&	$	14.1 (4.1)^{+1.3}_{-0.4}	$	&	$	11.4 (5.2)^{+1.2}_{-0.4}	$	\\ 
ln(${\rm lik}_{\rm WL}$)    	&	$	-81.3 (-79.3)^{+1.3}_{-0.5}	$	&	$	-	$	&	$	-82.6 (-80.5)^{+2.0}_{-1.1}	$	\\ 
ln(${\rm lik}_{\rm CMB}$)    	&	$	-	$	&	$	-294.3 (-290.1)^{+2.0}_{-1.1}	$	&	$	-295.5 (-290.8)^{+2.5}_{-1.3}	$	\\ 
$\chi^2_{\rm min}$	&	$	158.6	$	&	$	580.2	$	&	$	742.6	$	\\ 
    \hline
    \end{tabular}
    \caption{
    Same as Tab.~\ref{tab:lcdm_mcmc_results} but assuming the $\Lambda$DDM model.
    }
    \label{tab:dcdm_mcmc_results}
\end{table*}

\end{document}